\documentclass[twocolumn,graphics,floatfix,a4paper,prb]{revtex4-1}
\usepackage{color}
\usepackage[utf8]{inputenc}
\usepackage{graphicx}
\usepackage{amsmath}
\usepackage{amssymb}
\usepackage{longtable}

\usepackage{hyperref}
\hypersetup{
pdfnewwindow=true, colorlinks=true,
linkcolor=blue, anchorcolor=blue,
citecolor=blue, filecolor=blue,
menucolor=blue, urlcolor=blue}

\newcommand{\comment}[1]{\textit{}}
\newcommand{\bit}{\begin{itemize} \setlength{\itemsep}{0ex} \setlength{\topsep}{0ex} } 
\newcommand{\eit}{\end{itemize}}
\newcommand{\be}{\begin{equation}}
\newcommand{\ee}{\end{equation}}
\newcommand{\bea}{\begin{eqnarray}}
\newcommand{\eea}{\end{eqnarray}}
\newcommand{\ba}{\begin{align}}
\newcommand{\ea}{\end{align}}
\newcommand{\SKIP}[1]{}

\providecommand{\lr}{\ensuremath{\lambda_\mathrm{R}}}
\providecommand{\lnu}{\ensuremath{\lambda_\nu}}
\providecommand{\lso}{\ensuremath{\lambda_\mathrm{SO}}}
\providecommand{\zz}{\ensuremath{\mathbb{Z}_2}}

\providecommand{\bise}{B\lowercase{i}$_2$S\lowercase{e}$_3$}
\providecommand{\htr}{\ensuremath{H^\mathrm{SB}}}
\providecommand{\A}{\ensuremath{\mathcal A}}
\newcommand{\mb}{\ensuremath{\mathbf}}

\def \bk{{\bf k}}

\def \ua{\uparrow}
\def \da{\downarrow}

\begin{document}

\title{Smooth gauge and Wannier functions for topological band structures in arbitrary dimensions}
\author{Georg W. Winkler}
\affiliation{Institute for Theoretical Physics and Station Q, ETH Zurich, 8093 Zurich, Switzerland}
\author{Alexey A. Soluyanov}
\affiliation{Institute for Theoretical Physics and Station Q, ETH Zurich, 8093 Zurich, Switzerland}
\author{Matthias Troyer}
\affiliation{Institute for Theoretical Physics and Station Q, ETH Zurich, 8093 Zurich, Switzerland}

\date{\today}

\begin{abstract}
  The construction of exponentially localized Wannier functions for a set of bands requires a choice of Bloch-like functions that span the space of the bands in question, and are smooth and periodic functions of ${\bf k}$ in the entire Brillouin zone.  For bands with nontrivial topology, such smooth Bloch functions can only be chosen such that they do not respect the symmetries that protect the topology. This symmetry breaking is a necessary, but not sufficient condition for smoothness, and, in general, finding smooth Bloch functions for topological bands is a complicated task. We present a generic technique for finding smooth Bloch functions and constructing exponentially localized Wannier functions in the presence of nontrivial topology, given that the net Chern number of the bands in question vanishes. The technique is verified against known results in the Kane-Mele model. It is then applied to the topological insulator Bi$_2$Se$_3$, where the topological state is protected by two symmetries: time-reversal and inversion. The resultant exponentially localized Wannier functions break both these symmetries. Finally,  we illustrate how the calculation of the Chern-Simons orbital magnetoelectric response is facilitated by the proposed smooth gauge construction.
\end{abstract}

\maketitle
\section{Introduction}
\label{sec:intro}
Discoveries of the past decade established that not only the geometry of electronic bands but also their topology is a fundamental material property.\cite{Hasan_TI,Qi_and_Zhang_review} Being at the root of a variety of physical effects,~\cite{Qi_topological_2008,fu_majorana_2008} topologically nontrivial bands turned out to be ubiquitous, and many materials were found to host various topological states.\cite{molenkamp_hgte_2007,hasan_nat_2009,hasan_prl_2009,hasan_nat_2008,snte_exp}

As a general principle, nontrivial topologies in band structures are a consequence of some symmetry. They are examples of symmetry protected topological states.\cite{Wen_2D_2011,wen_symmetry_2013} For instance, magnetic insulators in 2D realize the integer quantum Hall effect in the absence of external magnetic field.\cite{haldane_model_1988,Volovik_He3_1988} In such insulators, dubbed Chern insulators, the nontrivial topology is captured by a Chern number topological invariant,~\cite{TKNN} which cannot be changed unless the $\mathrm{U}(1)$ charge conservation symmetry is broken.\cite{zhang_chargeconservation} In the case of time-reversal (TR) symmetric insulators, a $\mathbb{Z}_2$ topological invariant is assigned~\cite{KM_PRL_2005} to the band structure, that can not be changed without breaking the TR-symmetry (provided the band gap remains open). This concept of symmetry protection can also be generalized to crystalline symmetries, which protect the value of certain topological invariants, associated with these symmetries,~\cite{fu_tci_2011,bernevig_inversion} giving rise to crystalline topological insulators.\cite{snte_pred,snte_exp,ando2015topological}

The topology of a band structure in all these cases is associated with the Bloch bands and their Berry curvature~\cite{berry_1984, zak_berry_1989} in momentum space. An alternative description of crystalline solids can be made in position space by means of localized Wannier functions (WFs).\cite{wannier_1937, marzari_review_2012} Such a local basis is often preferable to that of Bloch functions, for example for modeling finite size effects in materials,~\cite{Souza-PRB01} chemical bonding,~\cite{Marzari_maximally_1997} computing electronic polarization,~\cite{King-Smith-PRB93} or ballistic transport.\cite{marzari_transport_2005} This variety of applications and the existence of established numerical techniques for WF-based analysis of materials,~\cite{wannier90_2008, marzari_review_2012} makes it important to extend the phenomenology of WFs to materials with nontrivial topology.

Another motivation for finding Wannier representation of topological bands is the calculation of the isotropic contribution to the orbital magnetoelectric response.\cite{Coh_Chern-Simons_2011} The linear magnetoelectric coupling tensor $\alpha_{ij}$ is defined as
\begin{equation}
  \alpha_{ij} = \left( \frac{\partial P_i}{\partial B_j} \right)_{\mathbf E = 0 } =
  \left( \frac{\partial M_j}{\partial E_i} \right)_{\mathbf B=0},
\end{equation}
where ${\bf P}$ and ${\bf M}$ are the polarization and magnetization of a material, ${\bf E}$ and ${\bf B}$ are electric and magnetic fields and both derivatives are evaluated at zero fields. While lattice and spin degrees of freedom also contribute~\cite{magnetoelectric_lattice,magnetoelectric_spin} to $\alpha_{ij}$ here we focus solely on the orbital (frozen-lattice) part. The orbital magnetoelectric response can be further split into two parts~\cite{Essin_magnetoelectric_2009,essin_magnetoelectric_2010}
\begin{equation}
  \alpha_{ij} = \tilde \alpha_{ij} + \frac{\theta e^2}{2 \pi h} \delta_{ij},
\end{equation}
of which $\tilde \alpha_{ij}$ is traceless, and the second, isotropic, part is characterized by a dimensionless quantity $\theta$, called the ``axion angle'' in high energy physics.\cite{wilczek_axion,Qi_topological_2008} There are two contributions to $\theta$:\cite{essin_magnetoelectric_2010} one is an ordinary perturbative Kubo term and the other is a purely geometrical one, the Chern-Simons contribution $\theta_\mathrm{CS}$. The latter one can be evaluated from the ground-state electron wave functions, by computing an integral of the Chern-Simons 3-form over the entire Brillouin zone (BZ).\cite{Qi_topological_2008,Essin_magnetoelectric_2009,essin_magnetoelectric_2010} However, this evaluation requires a choice of Bloch states that are smooth and periodic in $\mathbf k$-space as detailed in Sec.~\ref{sec:bise_magneto}. Alternative formulations of this term in position space require the existence of exponentially localized WFs (ELWFs).\cite{Coh_Chern-Simons_2011}

Magnetoelectric response was long thought to be observable only in materials that break TR and inversion symmetries. Remarkably, it turned out that TR- and inversion-symmetric topological insulators (TIs) are characterized by a nonzero quantized magnetoelectric response.\cite{Qi_topological_2008,Essin_magnetoelectric_2009,bernevig_inversion,turner_inversion_2012} In the presence of one of these symmetries the Kubo contribution to $\theta$ vanishes and $\theta=\theta_\mathrm{CS}$. The seeming contradiction is resolved by noticing that $\theta$ couples to $\mathbf E \cdot \mathbf B$ term in the Lagrangian, so that the corresponding equations of motion are invariant under $\theta \rightarrow \theta + 2\pi$.\cite{wilczek_axion} Since, in addition, $\theta$ is odd under both TR and inversion,~\cite{wilczek_axion} the {\it two} values $\theta = 0, \pi$ are compatible with these symmetries.

In TR- or inversion-symmetric TIs these two values of $\theta$ can be used as an analogue of the $\mathbb{Z}_2$ topological invariant,~\cite{Qi_topological_2008,Essin_magnetoelectric_2009} with $\theta=\pi$ corresponding to the TI phase and $\theta=0$ to the normal insulator (NI) one. For the cases of TR- and inversion-symmetric insulators the quantization of $\theta$ is exact and since $\theta$ corresponds to the $\mathbb{Z}_2$ invariant, it can be obtained using the methods for computing this invariant.\cite{Fu_time_2006,alexey_invariants_2011, yu_z2_2011} However, for materials that lack these symmetries, $\theta$ is in general not quantized, and it becomes necessary to evaluate the $\theta_\mathrm{CS}$-term directly. As mentioned above, this requires ELWFs. Since nontrivial band topologies can arise even when there are no symmetries that quantize the magnetoelectric response, computation of $\theta_\mathrm{CS}$ gives yet another motivation for developing a method to obtain ELWFs for topologically nontrivial band structures.

WFs are constructed by Fourier transforming Bloch states, and thus the momentum space geometry of the Bloch function can strongly influence the properties of the resultant WFs, in particular the degree of its localization. An important example is that the construction of an ELWF for a Bloch state with nonzero Chern number is impossible.\cite{Thouless_wannier_1984}  This is a consequence of the fact that a Chern number represents an obstruction for choosing the Bloch state to be a smooth and periodic function of ${\bf k}$ globally in the whole BZ.\cite{Panati-PRL07}

ELWFs can only be constructed if all the Bloch states, for which a Wannier representation is constructed, are smooth and periodic in the whole BZ. For topologically trivial bands, the choice of smooth Bloch states, referred to as a smooth gauge choice, respecting all the symmetries of the underlying band structure is generally possible. For topological bands, however, the symmetry that protects the topology represents an obstruction to choosing a smooth gauge that respects this symmetry. Consequently, the smooth WFs have to to break this symmetry. This was explicitly illustrated to be the case for the time-reversal (TR) symmetric ${\mathbb Z}_2$ TIs.\cite{Fu_time_2006,loring_disordered_2010,soluyanov_wannier_2011,roy_gauge_2009} \cite{soluyanov_wannier_2011}

While it is clear that the construction of a smooth gauge on a lattice requires breaking certain symmetries in the gauge, finding an explicit representation of smooth Bloch states  on a lattice of ${\bf k}$-points, required for a numerical construction of ELWFs, is a nontrivial task. An explicit construction based on parallel transport of the occupied states was obtained in Ref.~\onlinecite{alexey_gauge_2012} for a 2D model of a quantum spin Hall insulator. That construction, however, is tedious to generalize to the many-band case, and especially to higher dimensions.

Another approach to finding a smooth gauge is based on projecting certain localized orbitals that break the topology-protecting symmetry onto the the occupied states (see Sec.~\ref{sec:wf} for details) and is more appropriate for material calculations.\cite{soluyanov_wannier_2011} The problem of this method is that no specific algorithm for choosing the orbitals for the projection is presented. The requirement of breaking the topology-protecting symmetry in the initial projection is necessary but not sufficient for finding ELWFs.

In this work we develop a generic algorithm to construct ELWFs for a set of topologically nontrivial bands, with a zero net Chern number. The idea is to construct an adiabatic connection between trivial and TI phases by breaking the symmetries that protect the topology everywhere along the connecting path, except the initial and final points. Since the topology-protecting symmetries are broken at intermediate steps, it becomes possible to find a path connecting these two phases in a parameter space such that the insulating gap remains open along the connection.

The construction of this path allows one to avoid the problem of finding the correct symmetry-breaking projection in the topological phase. Instead, one discretizes the path in parameter space into steps, and constructs ELWFs at each step by projecting onto ELWFs found at the previous step. The initial step requires finding ELWFs in a topologically trivial case, which is a standard task.\cite{Marzari_maximally_1997} As a result, one naturally obtains ELFWs for the topological bands at the final point of the discretized path.

The breaking of symmetries plays a key role in adiabatically connecting two topologies. It is often the case that there are more than one symmetry protecting the topology of the band structure, and it is important that all of them are broken along the adiabatic path. For example, the ${\mathbb Z}_2$ topology of the TR-symmetric TIs can be additionally protected by inversion,~\cite{bernevig_inversion,alexandradinata_inversion_2014} in-plane mirror~\cite{fu_mirror,chiu_mirror} and certain other point group symmetries.\cite{bernevig_invariants} These symmetries need to be broken in addition to TR along the adiabatic path to obtain a smooth gauge in such band structures. 


This paper is organized as follows. In Sec.~\ref{sec:wf} we provide the theoretical background about construction of ELWFs. In this light, we also discuss the topological obstruction for constructing ELWFs in TIs. In Sec.~\ref{sec:avoid} In Sec.~\ref{sec:km} we apply our technique on the model of Kane and Mele~\cite{KM_PRL_2005}. Then in Sec.~\ref{sec:bise} we apply our technique onto \bise{} and use it to calculate the Chern-Simons magnetoelectric coupling $\theta_\mathrm{CS}$. Finally, we summarize our findings and give an outlook in Sec~\ref{sec:concl}.

\section{Wannier functions and topological obstruction}
\label{sec:wf}
%

%
\subsection{Construction of Wannier functions}
Here we provide a brief review of the methods introduced in the work of Ref.~\onlinecite{Marzari_maximally_1997} used for an explicit construction of ELWFs starting from a set of Bloch states obtained by diagonalizing the Hamiltonian on a mesh of ${\bf k}$-points. The  construction requires a projection of a set of trial localized orbitals onto the Bloch states. The choice of these orbitals, which is simple in the case of topologically trivial bands, becomes problematic for the case of nontrivial topology. For the method we present in this work, localized trial orbitals need to be found only in the topologically trivial case.

Given a Bloch state $\psi_{n{\bf k}}({\bf r})$, a corresponding WF is defined as
\begin{equation}
  \langle \mb r | \mb R n \rangle = W_n(\mb r - \mb R) = \frac{V}{(2\pi)^d} \int_\mathrm{BZ} d\mb k \, e^{-i \mb k \cdot \mb R} \psi_{n\mb k}(\mb r),
\label{eq:WF}
\end{equation}
where $V$ is the volume of the unit cell, $d$ is the dimensionality. The Bloch wave functions $\psi_{n\mb k}$ are assumed to be normalized within the unit cell. 

This definition, however, is not unique. The non-uniqueness, referred to as gauge freedom, is easily seen when constructing WFs for a set of $N$ Bloch states. A general unitary transformation $\mathrm{U}(N)$ of the $N$ occupied bands
\begin{equation}
|\tilde \psi_{n\mb k} \rangle = \sum_m U_{mn}(\mb k) | \psi_{m \mb k} \rangle ,
\label{trans}
\end{equation}
results in a different set of Bloch states that span the same Hilbert space as the original ones, and hence can equally be used for constructing the Wannier representation. Depending on the particular gauge choice, the resultant WFs and their degree of localization can vary a lot. In order to obtain ELWFs, the gauge choice has to be smooth, meaning that all $N$ Bloch states used to construct the WFs are smooth in the entire BZ, and obey periodic boundary conditions $\psi_{n{\bf k}}=\psi_{n{\bf k}+{\bf G}}$ upon translation by any reciprocal lattice vector ${\bf G}$. 

But even such a smooth gauge choice is not unique, since a smooth gauge transformation performed on a set of smooth Bloch states will result in a different set of smooth Bloch states and different ELWFs. Additional constraints can be put on the gauge to reduce the gauge freedom. A very common choice of such a constraint is the requirement of maximal localization of the resultant WFs in position space proposed in the work of Ref.~\onlinecite{Marzari_maximally_1997}. This gauge is obtained by minimizing the spread functional
\begin{equation}
\Omega=\sum_{n=1}^N \left( \langle \mb r^2\rangle_n - \langle \mb r \rangle_n^2\right)=\Omega_\mathrm{I}+\tilde{\Omega} ,
\end{equation}
where $\langle r\rangle_n=\int |W_n|^2 r d{\bf r}$, and $\Omega_I$ and $\tilde{\Omega}$ stand for the gauge-independent 
\begin{equation}
\Omega_\mathrm{I} = \sum_{n=1}^N \left[  \langle \mb r^2\rangle_n - \sum_{\mb R m} |\langle \mb R m | \mb r | \mb 0 n \rangle|^2 \right]
\end{equation}
and gauge-dependent parts of the spread
\begin{equation}
\tilde{\Omega} = \sum_{n=1}^N \sum_{\mb R m \ne \mb 0 n} |\langle \mb R m | \mb r | \mb 0 n \rangle|^2.
\end{equation}
The goal of maximal localization is to find a $\mathrm{U}(N)$ transformation that, when applied to some initial set of Bloch states according to Eq.~(\ref{trans}), minimizes $\tilde{\Omega}$ to produce a set of maximally localized WFs. Maximal localization can always be performed, once the initial choice of Bloch states is smooth.  
    
The necessity for smoothness of the initial Bloch states is most easily seen from the general procedure used to construct a set of WFs, as described in Ref.~\onlinecite{Marzari_maximally_1997}. Following this procedure, to construct a set of $N$ WFs from $N$ isolated (that is, separated by energy gaps from the rest of the spectrum) Bloch bands obtained from numerical diagonalization of the Hamiltonian, a set of $N$ localized trial states $|\tau_i\rangle$ is chosen. For each momentum $\mb k$ all of the trial orbitals are projected on the Bloch states to get a set of $N$ Bloch-like states~\cite{Marzari_maximally_1997}
\begin{equation}
  |\Upsilon_{i \mb k} \rangle = \hat P_{\mb k} |\tau_i \rangle = \sum_{n=1}^N |\psi_{n \mb k} \rangle \langle \psi_{n\mb k} | \tau_i \rangle ,
  \label{eq:proj}
\end{equation}
which are not orthonormal. 

For the construction of WFs these states need to be orthonormalized, and it is the orthonormalization procedure, where the smoothness of the gauge becomes crucial. To see this, apply Löwdin orthonormalization procedure to the states, which is commonly used  to get a set of orthonormalized Bloch states, to orthonormalize the states $|\Upsilon_{i \mb k} \rangle$
\begin{equation}
  |\tilde \psi_{n\mb k} \rangle = \sum_m [S(\mb k)^{-1/2}]_{mn} |\Upsilon_{m\mb k} \rangle ,
    \label{eq:S}
\end{equation}
where
\begin{equation}
S_{mn}(\mb k) = \langle \Upsilon_{m \mb k} | \Upsilon_{n \mb k} \rangle
\end{equation}
is the overlap matrix. While $|\tilde \psi_{n\mb k} \rangle$ are not the eigenstates of the single particle Hamiltonian, they span the same space as the usual Bloch eigenstates, and therefore describe the same ground state. For the trial states embodying a reasonable assumption about the character of the bands described, the $|\tilde \psi_{n\mb k} \rangle$ will be smooth functions of $\mb k$. In this case the WFs constructed from them by means of Eq.~\eqref{eq:WF} are expected to be exponentially localized, and the degree of localization is further increased by doing maximal localization. 

However, the orthonormalization procedure breaks down if at some ${\bf k}$ the determinant of the overlap matrix vanishes, that is $\det S({\bf k})=0$. This is guaranteed to happen if one runs into a topological obstruction, for example, by projecting onto a set of trial states $|\tau_i \rangle$ that respect the topology-protecting symmetry, as shown below.

\subsection{Topological obstruction via hybrid Wannier functions}
\label{sec:hwf}
\begin{figure}
\includegraphics[width =  0.5 \textwidth]{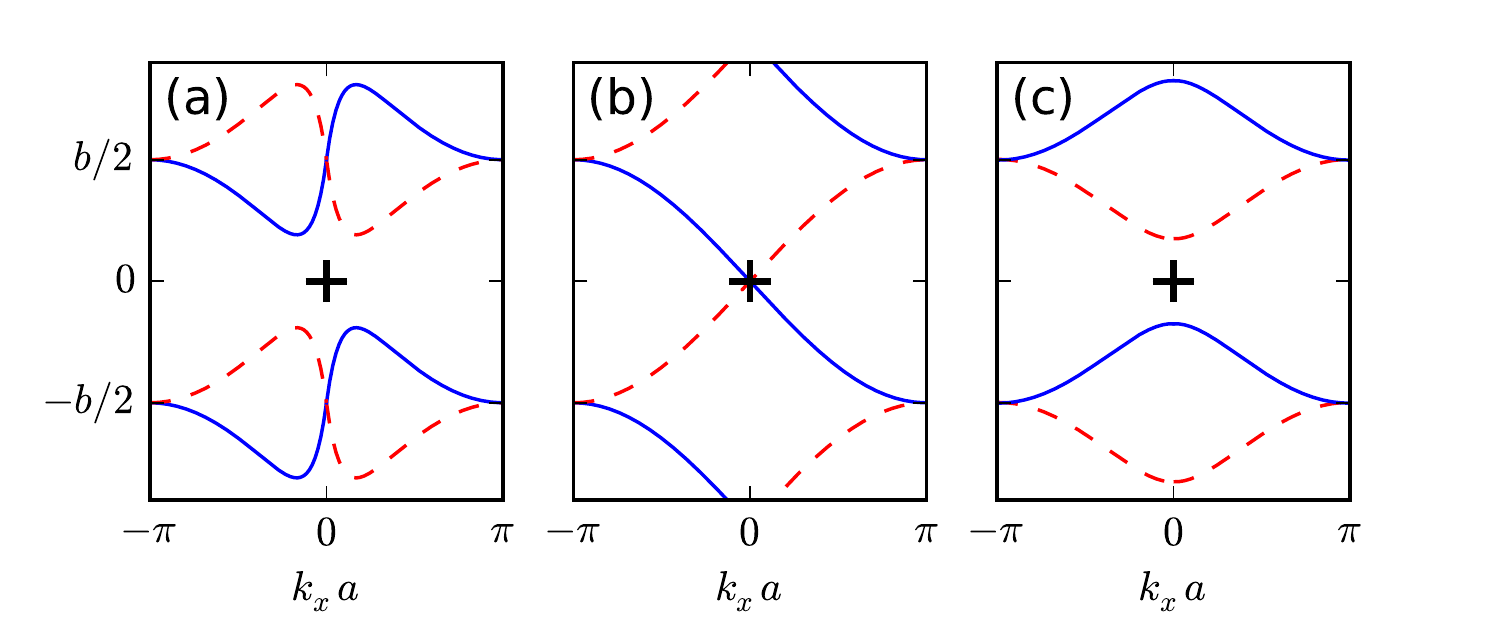}
  \caption{Possible flows of the hybrid Wannier charge centers in a 2D TR-symmetric system. Panel (a): topologically trivial insulator. Panels (b) and (c): quantum spin Hall insulator. A TR-symmetric gauge is used in panels (a) and (b), while a TR-breaking gauge is used in panel (c).}
  \label{fig:trpol}  
\end{figure}
Hybrid WFs~\cite{resta_hybrid_2001} (HWFs) differ from usual WFs in that the Wannier decomposition is done in one direction only
\begin{equation}
|R_y k_x n \rangle = \frac{a}{2 \pi} \int_{-\pi/a}^{\pi/a} dk_y e^{-iR_y k_y} |\tilde \psi_{n \mathbf k} \rangle .
\end{equation}
This wave function is localized in momentum in the $x$-direction and in position in the $y$-direction, being a hybrid of Bloch- and Wannier-like functions. HWFs proved to be useful for classifying topologies of band structures,~\cite{alexey_invariants_2011, gresch} providing an intuitive approach to topological invariants.

The band structure topology can be obtained by tracking the charge centers of HWFs defined as
\begin{equation}
\bar y_n(k_x) = \langle 0 k_x n | \hat y | 0 k_x n \rangle ,
\end{equation}
where the HWFs are located within the home unit cell. Computation of the charge centers does not require an explicit construction of HWFs, and can be done by means of the parallel transport procedure, as described in Ref.~\onlinecite{alexey_invariants_2011}. The HWF charge centers obtained in this gauge~\footnote{A discussion why parallel transport gives a valid gauge can be found in Sec. II B of Ref.~\onlinecite{alexey_invariants_2011}} are the eigenvalues of the projected position operator.\cite{kivelson_wannier_1982} However, for an isolated set of $N$ bands the gauge can be chosen differently, resulting in different values of $\bar{y}(k_x)$, and it is only the sum of all the $N$ centers that is gauge invariant (modulo a lattice vector in the $y$-direction) at each $k_x$.\cite{alexey_invariants_2011}

The parallel transport gauge respects the symmetries of the system. For example, in the presence of TR-symmetry the HWF charge centers  come in Kramers pairs $y_n(k_x)=y_m(-k_x)$ being doubly degenerate at the TR-invariant momenta $-k^*_x=k^*_x+G_x$, where $G_x$ is a reciprocal lattice vector in the $x$-direction. This is illustrated in  Fig.~\ref{fig:trpol}(a) and (b) for a model with two occupied bands.
The two centers are degenerate at $k_x=0$ and $k_x=\pm \pi/a$. Both centers evolve smoothly in between $\pm \pi/a$. In the case of Fig.~\ref{fig:trpol}(a) they return to the original value at the BZ boundary and the band structure is topologically trivial. In the case of Fig.~\ref{fig:trpol}(b) they interchange, which is generally the case in a quantum spin Hall insulator.\cite{Fu_time_2006} 

This interchange results in a discontinuity of the hybrid Wannier charge center lines as a function $k_x$: the two centers are continuous functions of momentum for $k_x\in [-\pi/a,\pi/a]$, but the periodicity constraint is not satisfied, meaning that the center position does not necessarily return to the original value after a $2\pi/a$ change in momentum. This is equivalent to placing the topology-dictated gauge discontinuity on the boundary of the BZ.
In this gauge choice, the Wannier centers each correspond to a particular, {\it individual} Chern number, since the shift of the Wannier center at the boundary is still an integer number of unit cells. Hence, each of the hybrid Wannier functions can be understood as a well-defined function, in a sense that a Chern number can be assigned to each of them, despite the possible presence of degeneracies in the energy spectrum.  

For example, the two bands in the Kramers pair are degenerate at the TR-symmetric momenta, but the corresponding hybrid Wannier centers result in a splitting of the pair into two subspaces with well-defined Chern numbers. For a trivial insulator, illustrated in Fig.~\ref{fig:trpol}(a), the two centers correspond to zero Chern numbers. For the case of a quantum spin Hall insulator the hybrid Wannier centers shown in Fig.~\ref{fig:trpol}(b) indicate that in the chosen gauge the occupied space is split into two subspaces with Chern numbers equal $\pm 1$.~\footnote{How this split may be actually facilitated on a lattice is described in Ref.~\onlinecite{alexey_gauge_2012}} In both cases the two subspaces are TR-images of each other, but the nonzero Chern numbers of the two states  in the topological phase signal that the TR-symmetric gauge is not smooth on the whole BZ torus. The above argument is equally valid for more than two occupied bands, where each of the two subspace contains several bands and Chern numbers are then assigned to each subspace.

The Kramers degeneracy of the Wannier centers persists for any gauge, in which TR maps one state at ${\bf k}$ onto the other state at $-{\mathbf k}$.\cite{alexey_gauge_2012} For this reason TR-symmetry in the gauge represents an obstruction for it to be smooth and periodic in the BZ. Tracking the presence of this obstruction is the basis for the methods of computing topological invariants.\cite{Fu_time_2006, alexey_invariants_2011,yu_z2_2011} 

The above analysis, in accord with other methods,~\cite{loring_disordered_2010,roy_gauge_2009,Fu_time_2006} leads to the conclusion that in the $\zz$ TIs a smooth gauge, and hence ELWFs, have to break TR-symmetry. Breaking the symmetry in the gauge (but not in the Hamiltonian) lifts the Kramers degeneracy of Wannier centers at the TR invariant momenta, and can result in a smooth gauge~\cite{soluyanov_wannier_2011} like the one illustrated in Fig.~\ref{fig:trpol}(c). The Wannier centers are smooth and single-valued throughout the BZ. This means that the occupied subspace was split~\cite{alexey_gauge_2012} into two subspaces with zero Chern numbers, and hence the construction of the ELWFs becomes possible in this gauge.

The above line of reasoning can be easily generalized to band structures, where the nontrivial topology is protected by a symmetry different from TR, for instance a crystalline symmetry.\cite{bernevig_inversion,turner_inversion_2012,fu_tci_2011,alexandradinata_ti_2014} A particular example can be that of the crystalline TI SnTe,~\cite{snte_pred,snte_exp} where the mirror Chern numbers~\cite{fu_mirror} of $\pm 2$ can be defined on the \{110\} mirror plane in the BZ. Hence, the mirror symmetry should be broken in the gauge to obtain ELWFs for this material. Again, not any symmetry-breaking gauge would work, but only specific ones that provide a decomposition of the occupied space into states that are smooth in the BZ. 

\section{Avoiding topological obstruction}
\label{sec:avoid}
\begin{figure}
   \includegraphics[width =  0.35 \textwidth]{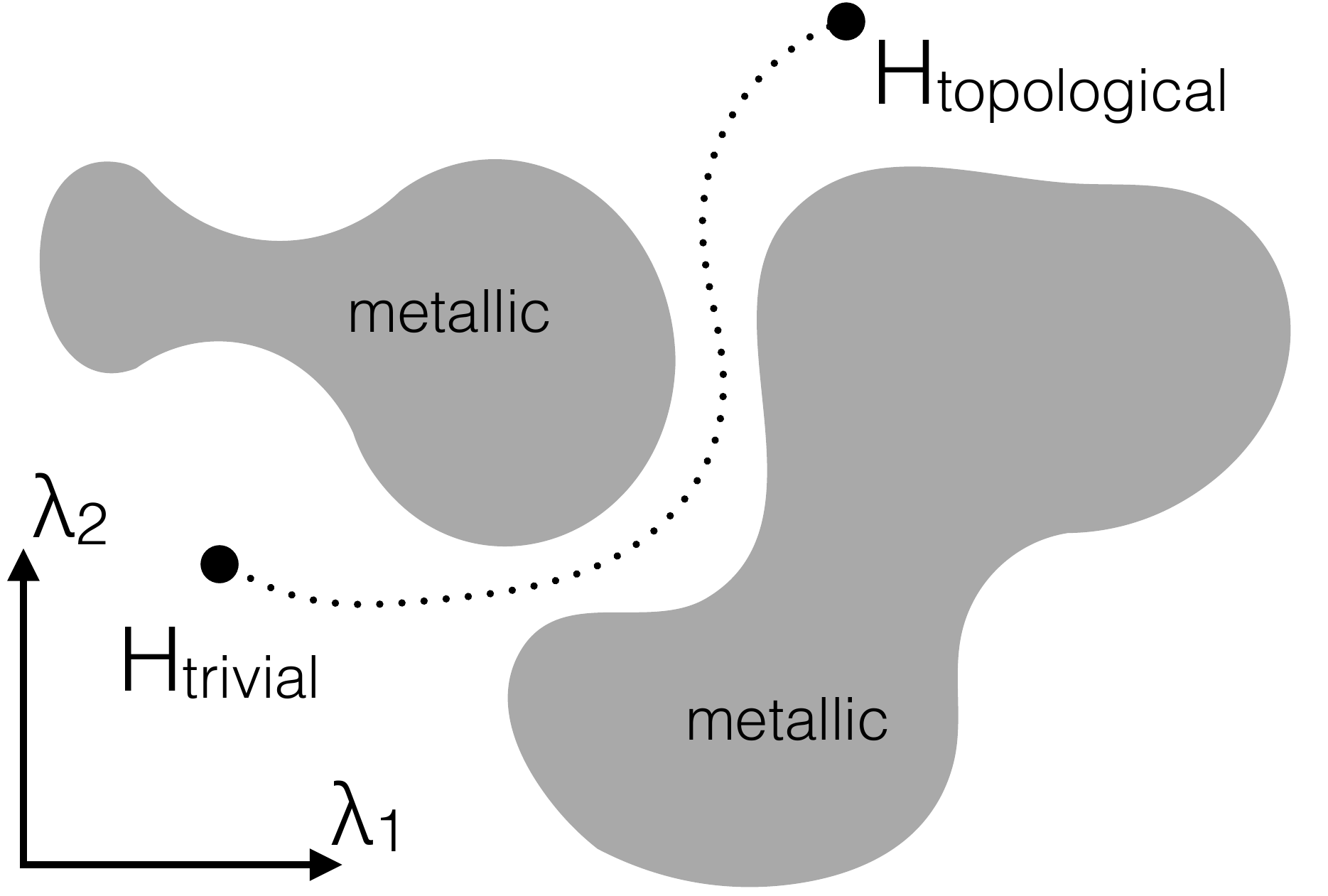}
  \caption{Schematic path in a parameter space of some model Hamiltonian. The adiabatic path is chosen to connect the topologically trivial and nontrivial phases avoiding gap closures. The topology-protecting symmetries is broken along the path.}
  \label{fig:symmetry_breaking}
\end{figure}
Finding suitable initial projections, for which $\det S(\mb k) \ne 0$ in the whole BZ, is particularly difficult for TIs due to the band inversion and hence, band character change, present in these materials. According to the discussion above and also the previous findings,~\cite{soluyanov_wannier_2011} the initial projections must be chosen such that they break the topology-protecting symmetries. However, as mentioned above, this requirement is necessary, but not sufficient.
Given that with an increasing number of bands the number of possible projections increases exponentially, a brute force search for suitable projections is complicated and  inefficient. 

Here we describe a technique for obtaining a smooth gauge and ELWFs for topological bands that avoids the need of finding suitable initial projections. Instead, we construct an adiabatic path that connects a topologically trivial Hamiltonian to the topological one, such that the system remains gapped along the path, as illustrated in Fig.~\ref{fig:symmetry_breaking}. To keep the system gapped, all the symmetries that protect the topology need to be broken along the path. The initial topologically trivial Hamiltonian can be chosen in different ways. For example, in  calculation for a material, where the nontrivial topology is often driven by spin-orbit coupling (SOC), the Hamiltonian with SOC artificially set to zero is topologically trivial and can be used as a starting point, if it is gapped.

Finding suitable trial functions $| \tau_i \rangle$ for topologically trivial bands is usually straightforward, since they respect the symmetries of the Hamiltonian.\cite{Marzari_maximally_1997} Moreover, an automated procedure for finding optimal projections for the trivial bands was proposed recently in Ref.~\onlinecite{coh_opf_2015}.
Thus, the ELWFs at the initial step of the path can always be found.

At least two parameters are required to parametrize the adiabatic path. One controls the topological phase transition by tuning the strength of symmetry breaking along the path.\footnote{The procedure is facilitated by choosing a path, along which the gap remains large on the scale of the topological gap. The larger is the gap, the coarser can be the discretization of the path.} Care should be taken when breaking the symmetries, since {\it double symmetry protection} can occur, meaning that there is more than one symmetry protecting the nontrivial topology. The issue of double symmetry protection is discussed with more detail in \ref{sec:double}.

Once the adiabatic path is constructed, the smooth Bloch states for the initial, topologically trivial, Hamiltonian $H_0$, are found and used to construct ELWFs. After this is done the path is discretized into $L$ steps, so that $H_L$ is the Hamiltonian of the TI in question. At step $1$ the Hamiltonian $H_1$ is diagonalized and the ELWFs obtained at the initial step are used as the trial orbitals for its occupied states. For dense enough discretization, the corresponding projection is guaranteed to have $\det{S}({\bf k})\neq 0$ throughout the BZ, since $H_1$ and $H_0$ are only slightly different. The resultant states are used to construct ELWFs for $H_1$. The procedure continues by projecting the ELWFs obtained at the step $\ell$ onto the occupied state of the Hamiltonian $H_{\ell+1}$, until the final point of the path is reached. Since at each step ELWFs were found, one ends up with ELWF-representation of the occupied topologically nontrivial space, which solves the problem of finding a smooth gauge.\footnote{We  also find it beneficial to run maximal localization of the WFs at each step of the procedure, prior to proceeding to the next one.} 

We emphasize that this method is general, being applicable to any isolated set of bands and in any dimension, provided that the net Chern number of this set is zero. 

\subsection{Double Symmetry Protection}
\label{sec:double}
Here we discuss the double symmetry protection of the topological phase using two examples: $\zz$ TR-symmetric insulators with mirror or inversion symmetries.

We start by considering the case of coexisting mirror and TR symmetries. When a TR-symmetric plane in the BZ of the insulator is invariant under mirror symmetry, the nontrivial 2D $\zz$ invariant of this plane suggests that both TR and mirror symmetries need to be broken in the smooth gauge. The mirror symmetry breaking in the gauge means that the Bloch states $|\tilde \psi_{n\bk}\rangle$ of Eq.~\ref{trans} obtained in the smooth gauge and used to construct ELWFs are not eigenfunctions of the mirror operator on the mirror plane. This can be seen by noting that the two  states in a mirror-symmetric Kramers pair have opposite mirror eigenvalues $\pm i$ on this plane, so the occupied space of an insulator on this plane can be split into two subspaces according to the value of the mirror eigenvalue. In the $\zz$-odd phase $+i$ and $-i$ have opposite odd Chern numbers, and thus breaking TR symmetry only is not sufficient to remove the topological obstruction, since the two mirror-labeled subspaces remain nontrivial. Hence, the mirror symmetry has to be broken as well, to construct Bloch states that are smooth in the entire BZ for the $\zz$-odd phase. 

Another example of double protection is that of an inversion- and TR-symmetric TI. The smooth gauge in this case also has to break both symmetries, as we argue below. Several works~\cite{turner_inversion_2012,alexandradinata_inversion_2014} discussed the possibility of a topological classification in the presence of inversion-symmetry only. In particular, the work of Ref.~\onlinecite{alexandradinata_inversion_2014} presented a study of the HWF centers in inversion symmetric insulators, obtained by a special construction (called a Wilson loop~\cite{yu_z2_2011}). In that construction the projector onto all the occupied states $\hat{P}^{\rm{o}}_{\bf k}$ is defined at each $k$-point, and the HWF centers are obtained by taking the log of the eigenvalues of the product of operators $\prod_\bk \hat{P}_{\bf k}^{\rm{o}}$ taken over discretized values of $k$ on a closed loop in momentum space~\cite{yu_z2_2011}. For inversion symmetric systems the inversion symmetry $\hat I$ puts the following constraint on the projector $\hat{P}_{\bf k}^{\rm{o}}=\hat{I}\hat{P}_{-{\bf k}}^{\rm{o}}\hat{I}$. 

The resultant HWFs are inversion symmetric in a sense that for each center $\bar{y}_n(k_x)$ there exists an inversion symmetric partner $-\bar{y}_n(-k_x)$, where $n$ is not necessarily equal to $m$. Thus, in the presence of inversion symmetry the sum of all the HWF centers at the inversion-invariant momenta $k_x=\{0,\pi/a\}$ can only be $0$ or $b/2$, which are the inversion-symmetric values assuming the inversion center coincides with the center of the unit cell.  

However, as discussed above, when constructing ELWFs, one needs a representation of the occupied subspace in terms of Bloch states that are smooth and periodic in $\bk$-space, meaning that the projector on each of these Bloch states is smooth. Thus, in a smooth gauge the net projector onto the occupied states is decomposed into a set of projectors $\hat{P}^{\rm{o}}_{\bf k}=\sum\hat P^{n}_{\bf k}$, each of which is smooth. However, imposing inversion symmetry constraint of the form $\hat{P}_{\bf k}^{n}=\hat{I}\hat{P}_{-{\bf k}}^{n}\hat{I}$ on each of the individual projectors, restricts the corresponding HWF centers (not their sum, but each of them separately) to take on inversion-symmetric values at $k_x=\{ 0,\pi/a\}$, that is either $0$ or $b/2$. We call a gauge, in which each of the projectors respects inversion symmetry in this sense an inversion-symmetric gauge.

In an inversion-symmetric gauge each Bloch state $|\tilde \psi_{n\bk} \rangle$ respects inversion symmetry $e^{i\phi}|\tilde \psi_{n\bk} \rangle = \hat I |\tilde \psi_{n-\bk} \rangle$, and hence the corresponding WF centers are subject to the condition $\langle {\bf r} \rangle_n = - \langle {\bf r} \rangle_n \mod {\bf R}$. Consequently, the HWF centers fulfill $\bar y_n(k_x) = -\bar y_n(-k_x)$ modulo $b$ and are restricted to the values $\bar y_n(k_x) = \{0,b/2\}$ at $k_x = \{0, \pi/a\}$. Note, that the HWFs are smooth in momentum in the interior of the BZ, and thus the index $n$ refers to the same center on both sides of this equation. If for $k^*_x=\{0,\pi/a\}$,  $|\tilde \psi_{n\bk} \rangle$ has the same parity at $k_y = 0$ and $\pi/b$ then $\bar y_n(k_x^*)=0$, while if the parities are opposite $\bar y_n(k_x^*)=b/2$. 

This means that if an inversion-symmetric Bloch state has a non-zero Chern number on some 2D BZ (2D cut of a 3D BZ), thus being not smooth, it is impossible to make it smooth (that is, change the Chern number to zero) if the inversion symmetry in the gauge is preserved. This is shown in Fig.~\ref{fig:trpol}. The illustrated centers are obtained in an inversion-symmetric gauge, and by construction are smooth in the interior of the BZ $k_x\in [-\pi/a,\pi/a]$. The Chern numbers of the centers shown as solid blue and dashed red lines in Fig.~\ref{fig:trpol}(b) are given by the number of unit cells traversed by the HWF center when going from one edge of the BZ to the other, and are equal to $\pm 1$. They cannot be changed to zero by breaking TR-symmetry alone, while preserving inversion symmetry in the gauge in the above sense, since that would require moving the center position away from the value fixed by inversion symmetry at some high-symmetry momentum. In the inversion-symmetric gauge, where each Bloch state separately is taken to be the eigenstate of inversion operator, this argument holds for any number of Kramers pairs, since the argument applies for each individual HWF center.

To state it more rigorously, the TR and inversion invariant $\mathbb{Z}_2$ topological insulating state in 2D is characterized by the total number of negative parity eigenvalues of occupied states at the four TR invariant momenta:\cite{ti_with_inversion,alexandradinata_inversion_2014} if this number divided by two is odd the system is a topological insulator.\footnote{This argument is easily generalized to weak and strong 3D TIs, taking into account that these phases have certain ordering of trivial and non-trivial 2D TR-symmetric planes in the BZ.~\cite{fu_3d_2007}}  Let $n^{-}_1$ be the difference in the number of negative parity eigenvalues between points $(0,0)$ and $(0,\pi/b)$, and $n^{-}_2$ the difference between points $(\pi/a,0)$ and $(\pi/a,\pi/b)$. If $n^{-}_1$ and $n^{-}_2$ are different (which is the case in a TI) then the number of HWF centers for which $\bar y_n(k_x)=b/2$ is also different at $k_x=0$ and $k_x = \pi/a$. Since the HWF centers are smooth in the BZ interior $k_x \in [-\pi/a,\pi/a]$,  at least $|n^{-}_1 -n^{-}_2|$ of the hybrid centers $\bar y_n(k_x)$ correspond to a nonzero Chern number. These Chern numbers can only be removed by breaking inversion symmetry of individual $|\tilde \psi_{n\bk}\rangle$, similar to the TR symmetric case discussed in Sec.~\ref{sec:hwf}. Therefore, the obstruction to smoothness persists in the inversion symmetric gauge. Note that also in a $\zz$-even case inversion symmetry can protect additional topologies as has been pointed out in Ref.~\onlinecite{turner_inversion_2012,alexandradinata_inversion_2014}.

Thus, the adiabatic connection of an inversion-symmetric topological insulator to a trivial one should be found by breaking the inversion symmetry along the path.

\section{Application to Kane-Mele model}
\label{sec:km}
We first illustrate our technique by applying it to the Kane-Mele (KM) model that describes a 2D quantum spin Hall ($\zz$-odd) insulator in some of its parameter space.\cite{KM_PRL_2005}  A smooth gauge and the corresponding WFs were obtained previously for this model by other methods,~\cite{soluyanov_wannier_2011, alexey_gauge_2012} and it is instructive to validate our method versus known results before going to more complicated cases.
\subsection{Kane-Mele model}
The KM is a tight-binding model on a honeycomb lattice with one spinor orbital per site. The primitive hexagonal lattice vectors are $\boldsymbol{a}_1 = a \boldsymbol{\hat x}$ and $\boldsymbol{a}_2 = \frac a 2 (\boldsymbol{\hat x}+\frac{\sqrt{3}}{2} \boldsymbol{\hat y})$ with the atoms $A$ and $B$ located at at sites $\boldsymbol{t}_A = \frac 1 3 (\boldsymbol a_1 + \boldsymbol a_2)$ and $\boldsymbol{t}_B = \frac 2 3 (\boldsymbol a_1 + \boldsymbol a_2)$. The KM Hamiltonian is given by
\begin{align}
  \begin{split}
  H=&\sum_{<ij>}  t_{ij}\, c_i^\dag c_j + i \lso \sum_{\ll ij \gg} \nu_{ij} c^\dag_i \sigma_z c_j \\
   &+ i \lr \sum_{<ij>}c^\dag_i ({\bf s} \times \boldsymbol{\hat d}_{ij})_z c_j+
  \lnu \sum_i \xi_i c^\dag_i c_i ,
  \end{split}
  \label{eq:kmH}
\end{align}
where summation over the suppressed spin indices is assumed. The summation $<ij>$ runs over all nearest neighbors, and the sum over $\ll ij\gg$ runs over all second-nearest neighbors.  $\nu_{ij}=(2/\sqrt{3})(\boldsymbol{\hat d}_1 \times \boldsymbol{\hat d}_2)=\pm 1$, with $\boldsymbol{\hat d}_1$ and  $\boldsymbol{\hat d}_2$ being the first-neighbor bond vectors encountered by an electron hopping from $j$ to $i$, ${\bf s}$ is a spin-1/2 operator.  $\lso$ and $\lr$ are parameters defining the spin-orbit coupling and $\lnu$ is a staggered onsite potential. In what follows we fix $t=1$, $\lambda_{SO}=0.6$ and $\lambda_R = 0.5$.

\subsection{Constructing an obstruction-free path for the Kane-Mele model}
\begin{figure}
\includegraphics[width =  0.45 \textwidth]{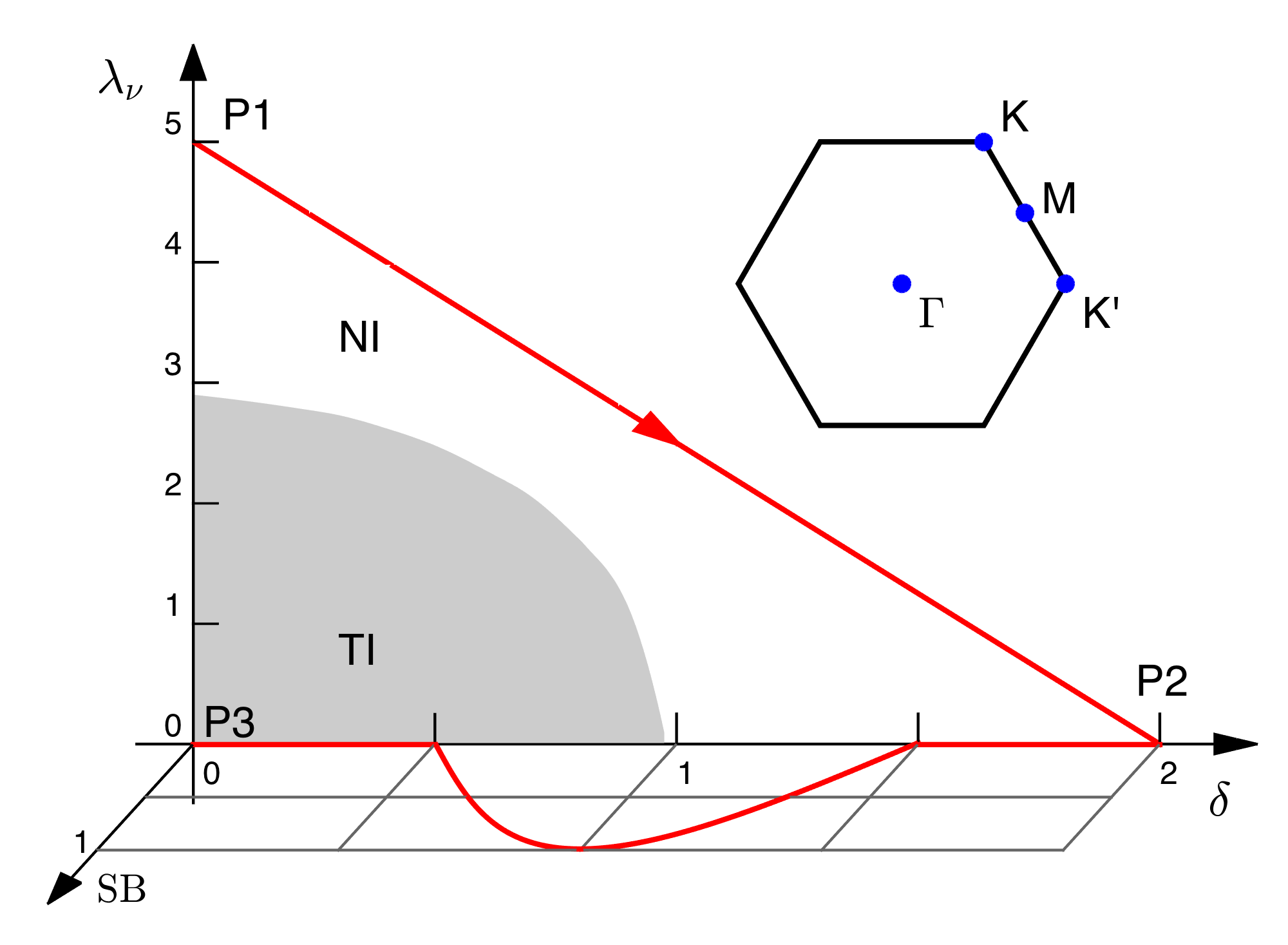}
  \caption{Phase diagram of the KM model for $\lso=0.6$ and $\lr = 0.5$. The red arrows indicate the path used to construct an obstruction-free gapped path between the trivial and topological phases. The symmetry breaking is indicated in the third, out of plane, axis. The BZ of a honeycomb lattice is shown in the inset.}
  \label{fig:km_pd}  
\end{figure}
Following our method, first an adiabatic path connecting the normal insulator (NI) to the TI phase of the KM model needs to be found. Let us first tune $\lnu$ while keeping the spin-orbit coupling parameters $\lso$ and $\lr$ fixed. This path corresponds to the vertical axis in Fig.~\ref{fig:km_pd}. A gap closure occurs at the two Dirac points at $K$ and $K'$ in the BZ (illustrated in the inset of Fig.~\ref{fig:km_pd}). We thus need to find a symmetry breaking field that would prevent the gap closure at both Dirac points simultaneously. 

We thus introduce a hopping anisotropy $\delta$ to the nearest-neighbor hopping $t_{ij}$. For the reasons that are made clear below, it is chosen such that the hopping is $t(1+\delta)$ in the $(1,1)$ direction and $t$ in the other directions, and it breaks the $C_3$-rotational symmetry of the KM model (see Appendix~\ref{appendix} for a thorough discussion of the symmetries in the KM model). By tuning $\delta$ and moving along the $\lnu=0$ line in Fig.~\ref{fig:km_pd} the gap closure can be moved to the $M$ point. This  path is depicted by the red arrows in Fig.~\ref{fig:km_pd}. 

The degeneracy at the $M$ point is easily lifted by an appropriate weak TR-symmetry breaking field. We found the optimal field to be a staggered magnetic field in the $(-1,1)$-direction of the form\footnote{We found that the field of the form 
$$
  \htr_{(1)} = (-\frac{1}{2} \sigma_x + \frac{\sqrt 3}{2}\sigma_y)\tau_z,
  \label{eq:hsb1}
$$
works as well, but it breaks less symmetries than $\htr_{(2)}$, and the resultant WFs are less localized compared to those obtained with $\htr_{(2)}$, as explained in the Appendix~\ref{appendix}.}
%
\begin{equation}
  \htr_{(2)} = (-\frac{\sqrt 3}{4} \sigma_x + \frac{3}{4}\sigma_y+\frac{1}{2} \sigma_z)\tau_z.
  \label{eq:hsb2}
\end{equation}
where $\sigma$ and $\tau$ are the Pauli matrices acting in the spin and sublattice subspaces correspondingly. Apart from the TR-symmetry, this field breaks the mirror and $C_2$-rotational symmetries of the KM model (see Appendix~\ref{appendix}) and the resultant WFs coincide with the ones reported in Ref.~\onlinecite{soluyanov_wannier_2011}. 

The final path, shown by the red line in Fig.~\ref{fig:km_pd}, is thus made of two parts: the line $\overline{\mathrm{P1}\,\mathrm{P2}}$, connecting two points with the same $\zz$ invariant, and the line $\overline{\mathrm{P2}\,\mathrm{P3}}$, where the $\zz$ invariant changes from even to odd, and along which the TR symmetry is broken. Note that it is also possible to start the path directly at P2, which corresponds to the NI phase. However, for reasons made clear below, we find it illustrative to add $\overline{\mathrm{P1}\,\mathrm{P2}}$ to the discussion.

\subsection{Wannier functions of the KM model}
\begin{figure}
\includegraphics[width =  0.45 \textwidth]{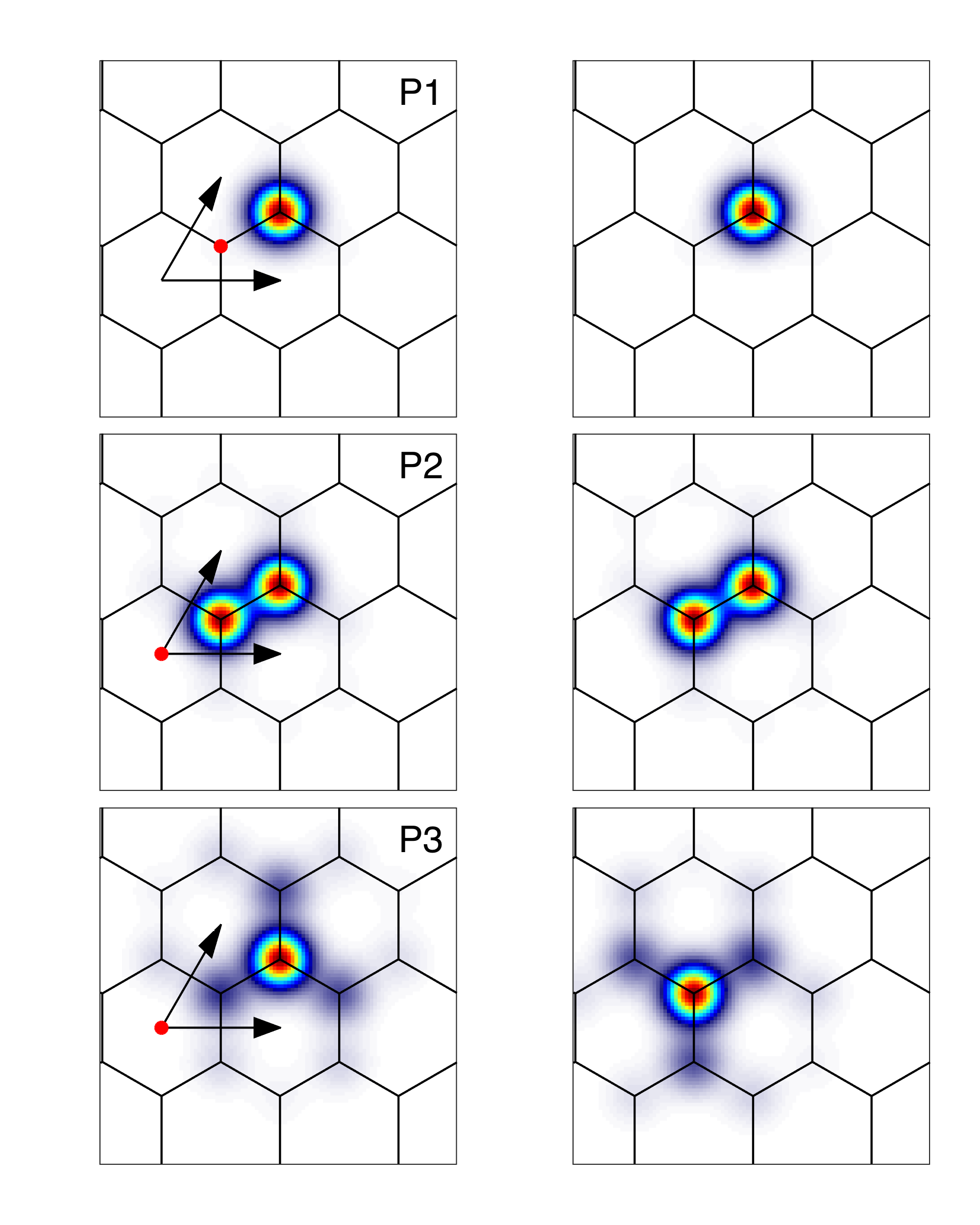}
\caption{Charge density distribution corresponding to the WFs obtained at points P1, P2 and P3 of the Fig.~\ref{fig:km_pd}. The left and right panels show the two WFs of the KM model at each of the points. The red point marks the sum of the corresponding Wannier centers (modulo a lattice vector).}
  \label{fig:km_wf}  
\end{figure}
Now that the gapped path connecting points P1 and P3 is found, the above outlined method is implemented and the WFs are calculated along this path. The position space density\footnote{Gaussian-shaped orbitals were used for the graphical representation.} of the WFs is shown in Fig.~\ref{fig:km_wf} corresponding to the points P1,P2 and P3 on the adiabatic path of Fig.~\ref{fig:km_pd}.

At P1 (NI phase) both WFs are localized on the lower energy sites. Each of them is mapped onto the other by TR, forming a Kramers pair of WFs with opposite spins. 

The sum of the Wannier centers $\boldsymbol{P}=\bar{\bf r}_1+\bar{\bf r}_2$, being proportional to the electronic polarization,~\cite{King-Smith-PRB93} is illustrated with a red dot in the left panel in Fig.~\ref{fig:km_wf}. The $C_3$-symmetry of the KM model (see Appendix~\ref{appendix}) constrains the possible values of this sum, to the values that are invariant under this symmetry modulo a lattice vector. Consistent with $C_3$ are the following values: $(P_x,P_y)=(0,0),(1/3,1/3),(2/3,2/3)$. A change between distinct values cannot occur continuously, unless the $C_3$ symmetry is broken. The value of $\boldsymbol{P}$ corresponding to the point P1 is $P_x=1/3$ and $P_y=1/3$. This is yet another example of additional topological protection. If there is a symmetry that quantizes the values of electronic polarization~\cite{King-Smith-PRB93}, and if the values in the topologically trivial and nontrivial phases are different, then the symmetry must be broken so that the polarization  can be  changed continuously between the two discrete values.\cite{top_polarization}

When going from P1 to P2, $\delta \neq 0$ and $\lnu \neq 0$ and both the $C_3$ and $C_2$-symmetries are broken. This allows to continuously change the electronic polarization  without closing the band gap. At P2 (NI phase) the two lattice sites have equal energies, since $\lnu = 0$. Accordingly, the two WFs are equally distributed between both sites, again forming a Kramers pair. Note that $P_x=P_y=0$ at this point.

At P3 we are in the desired $\zz$-odd phase. The field of Eq.~\ref{eq:hsb2} is introduced to get from P2 to P3 along the gapped path. The two WFs are localized on different sites, having opposite spins, in accord with the earlier study of Ref.~\onlinecite{soluyanov_wannier_2011}. Clearly, this configuration does not conserve the TR symmetry of the system. The sum of the Wannier centers remains at $P_x=P_y=0$. 
\begin{figure*}
  \includegraphics[width = 0.45 \textwidth]{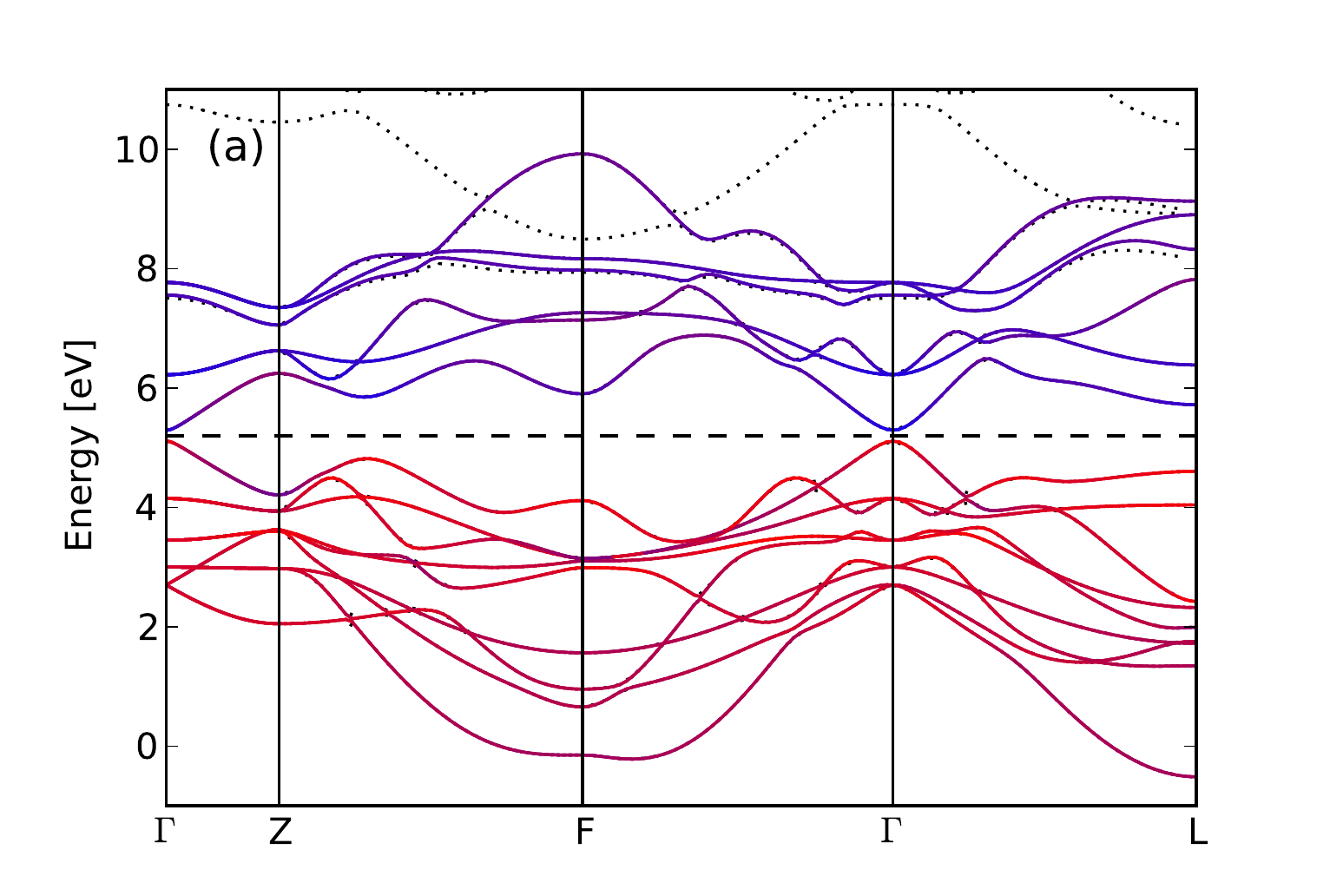}  \includegraphics[width = 0.45 \textwidth]{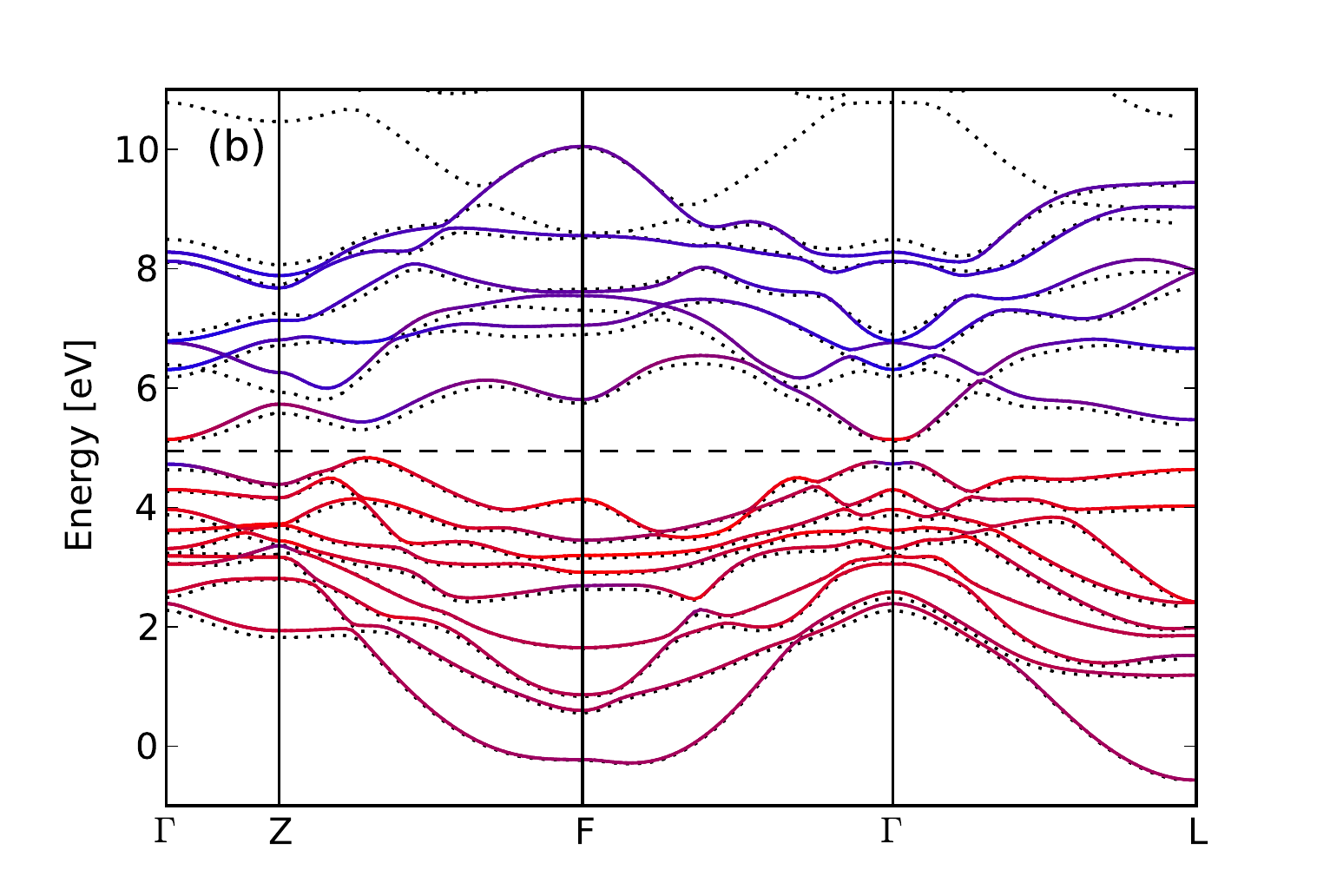}
  \caption{Projected band structure (Se (Bi) character shown in red (blue)) and the {\it ab initio} band structures (black dotted lines) of \bise{}.  Left panel: no spin-orbit coupling. Right panel: spin orbit coupling included.}
  \label{fig:bands}
\end{figure*}
\section{Application to \bise{}}
\label{sec:bise}
The standard example~\cite{hasan_nat_2009, Zhang_Nat_2009} of a TI is \bise{}. It has a rhombohedral lattice structure with the space group $D^{5}_{3d}$. The material is layered with hexagonal quintuple layers, consisting of three Se (two equivalent Se1 and one Se2) and two equivalent Bi atoms (shown as {\bf A}, {\bf B} and {\bf C} in Fig.~\ref{fig:crys_struct}) bounded together by van der Waals interaction. The structure is inversion-symmetric, with an inversion center at the central Se0 atom marked with the cross in Fig.~\ref{fig:crys_struct}. 

Without SOC the band structure of \bise{} is topologically trivial, turning SOC on with a parameter $\lso$ drives it into the TI phase. At $\lso=0$ \bise{} is a direct small band-gap semiconductor (NI phase), whereas at the full experimental SOC strength $\lso = 1$ it is a TI. At the topological phase transition there is an intermediate semi-metallic state with a gap closure. This gap closure can be avoided by applying a suitable symmetry breaking field.

\subsection{Constructing a model Hamiltonian for \bise{}}
A fully self-consistent density functional theory calculation was carried out for \bise{} without spin-orbit coupling (SOC) using the Vienna {\it ab initio} simulation package (VASP)~\cite{kresse_vasp_1996,kresse_vasp1_1996} with the projector augmented-wave method,\footnote{We used 340 eV for the cut-off energy and a 10$\times$10$\times$10 ${\bf k}$-point mesh in the BZ. The lattice constants were taken from Ref.\onlinecite{wyckoff}.} using generalized gradient approximation of the Perdew--Burke--Ernzerhof for the exchange-correlation potential.\cite{PBE_1996} The Wannier90~\cite{wannier90_2008} package was then used to first disentangle an isolated group of bands and then to project  the band structure onto the atomic $p$-orbitals of all Bi and Se atoms. No further iterative minimization of the WF spread was done.

The resultant band structures, obtained from the projected Hamiltonian and the {\it ab initio} calculations are illustrated in the left panel of Fig.~\ref{fig:bands}.  SOC is added afterwards to this Hamiltonian by adding a local Hamiltonian $H^\mathrm{SOC}$ in the basis of the atomic $p$-orbitals.\cite{Zhang_NJP_2010, chadi_spinorbit_1977} The case of full SOC is illustrated in the right panel of Fig.~\ref{fig:bands}. The correct topological phase of the projected Hamiltonian was confirmed using the method of Ref.~\onlinecite{alexey_invariants_2011}

In the following we work with the projected Hamiltonian, which allows for the easy control of SOC and a symmetry breaking field in the total Hamiltonian
\begin{equation}
  H^\mathrm{tot} = H^0 + \lso\, H^\mathrm{SOC} + \alpha\, \htr.
  \label{htot}
\end{equation}
The term $\htr$ is introduced below. Note that both, $H^\mathrm{SOC}$ and $\htr$ are local. The two parameters $\lso$ and $\alpha$ are found to be sufficient to construct an adiabatic path between the NI and TI phase in \bise{}. 

\subsection{Constructing an obstruction-free path for \bise{}}
\begin{figure}
  \includegraphics[width = 0.2 \textwidth]{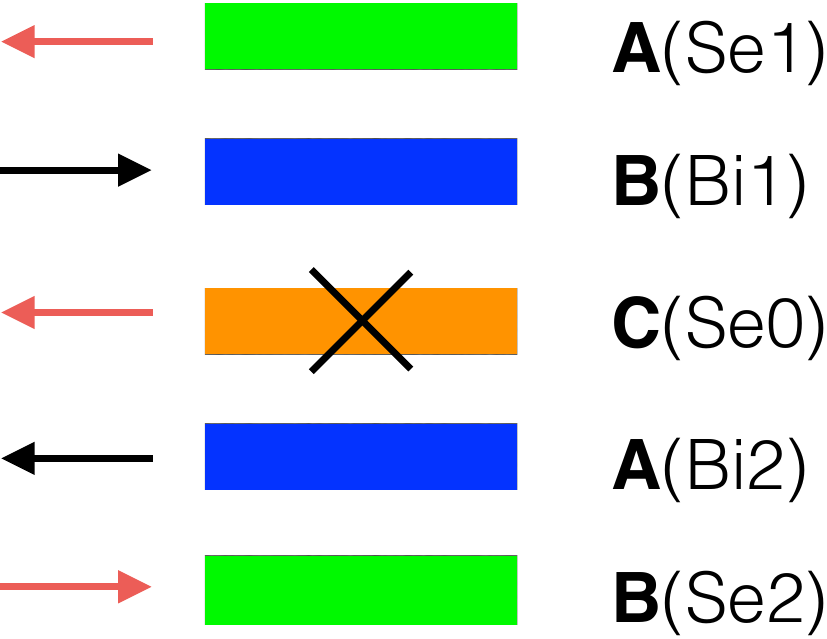}
  \caption{The layered structure of \bise{}. The cross denotes the inversion center at the Se0 atoms. The black arrows indicate the staggered magnetic field on the Bi sites corresponding to $\htr_\mathrm{Bi}$. The red arrows indicate an additional field corresponding to $\htr_\mathrm{BiSe}$.}
  \label{fig:crys_struct}
\end{figure}
The SOC strength $\lso$ is tuned from $\lso=0$ (NI phase) to $\lso=1$ (TI phase with full experimental SOC strength),\cite{spin-orbit} passing through a topological phase transition at $\lso \approx 0.47$. At this phase transition the gap closes and a 3D Dirac cone is formed at the $\Gamma$ point. To obtain a gapped adiabatic path connecting NI and TI phases, a suitable symmetry breaking field $\htr$ that gaps the Dirac cone needs to be found by considering the symmetry breaking requirements. Due to the double protection described in Sec.~\ref{sec:avoid}, the field must break both TR and inversion symmetries. This requirement is satisfied by a staggered magnetic field, analogous to that of Eq.~(\ref{eq:hsb2}) used in the case of the KM model. The optimal direction of the staggered field was found to be in the plane of hexagonal layers (see Fig.~\ref{fig:crys_struct}).

Examples of two such staggered fields are illustrated in Fig.~\ref{fig:crys_struct}. The first one, $\htr_\mathrm{Bi}$, shown with black arrows, acts on Bi sites only. While both TR and inversion symmetries are broken by this field, the combination of the two symmetries survives. As a consequence, the band structure remains doubly degenerate upon the inclusion of this field. 

The other symmetry breaking field, $\htr_\mathrm{BiSe}$, breaks this compound symmetry as well by applying, in addition to $\htr_\mathrm{Bi}$, a field on all Se sites, as marked by the red arrows in Fig.~\ref{fig:crys_struct}. Application of this field lifts the double degeneracy of the bands. Both fields prevent the gap closure and allow for the adiabatic connection between the NI and the TI.
\begin{figure}
  \includegraphics[width =  0.45 \textwidth]{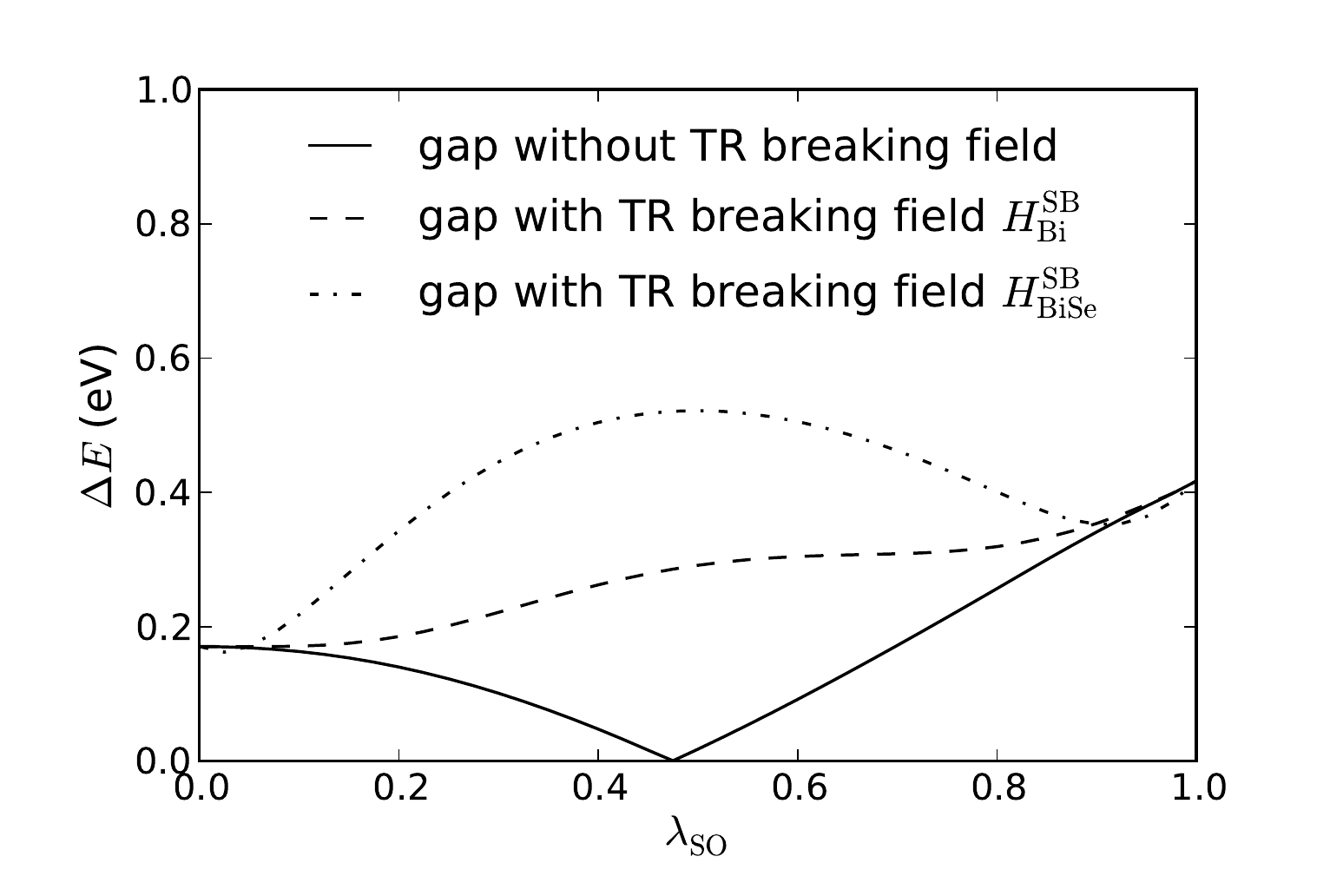}
  \caption{The bulk energy gap as a function of $\lso$. The symmetry breaking is tuned as $\sin(\lso \pi )$. The field $\htr_\mathrm{Bi}$ preserves the double degeneracy of bands. The gap is larger for the $\htr_\mathrm{BiSe}$ field, which lifts this degeneracy.
  }
  \label{fig:bisegap}
\end{figure}

The band gap as a function of $\lso$ is shown in Fig.~\ref{fig:bisegap}. The parameter $\alpha$ of the TR symmetry breaking field is tuned as $\alpha(\lso) = \sin(\lso  \pi)$.
We interpolate the path between $\lso=0$ and $\lso=1$ using 9 intermediate equidistant steps. In what follows the BZ of the Hamiltonian~\ref{htot} is discretized into a 16$\times$16$\times$16 ${\bf k}$-mesh. 
\subsection{Wannier functions of \bise{}}
\label{sec:bise_wf}
\begin{table}
\begin{tabular}{ l || c | c || c | c}
    & \multicolumn{2}{|c||}{$\lso =0$} & \multicolumn{2}{|c}{$\lso =1$} \\ \cline{2 - 5}
    & $\Omega_\mathrm{I}$ &  $\tilde \Omega$ & $\Omega_\mathrm{I}$ &  $\tilde \Omega$  \\ \hline
  $\htr_\mathrm{Bi}$& 11.01 & 0.46 & 11.04 & 0.70 \\
  $\htr_\mathrm{BiSe}$& 11.01 & 0.46  & 11.04 & 0.73
\end{tabular}
\caption{The spread of the resultant WFs after minimization.}
\label{tab:ws}
\end{table}
We first need to find WFs for the initial step of the path, that is for $\lso=0$. Suitable trial states $|\tau_i \rangle$ for the topologically trivial band structure can be guessed from the occupations of atoms shown with color in Fig.~\ref{fig:bands}(a). The occupied subspace consists mainly of the Se $p$-states. Therefore, the Se $p$-orbitals (both up and down spin) are chosen to be the initial trial states. 

Throughout the path we monitor the minimum of $\det{S({\bf k})}$ and the Wannier spreads. Apart from the first projection, where we start with local projections, the determinant always stays reasonably close to $1$. For this reason, the minimization of the Wannier spreads done at each step of the path, after projecting onto the WFs obtained at the previous step, results in only small improvement in localization.

\begin{figure}
\includegraphics[width =  0.5 \textwidth]{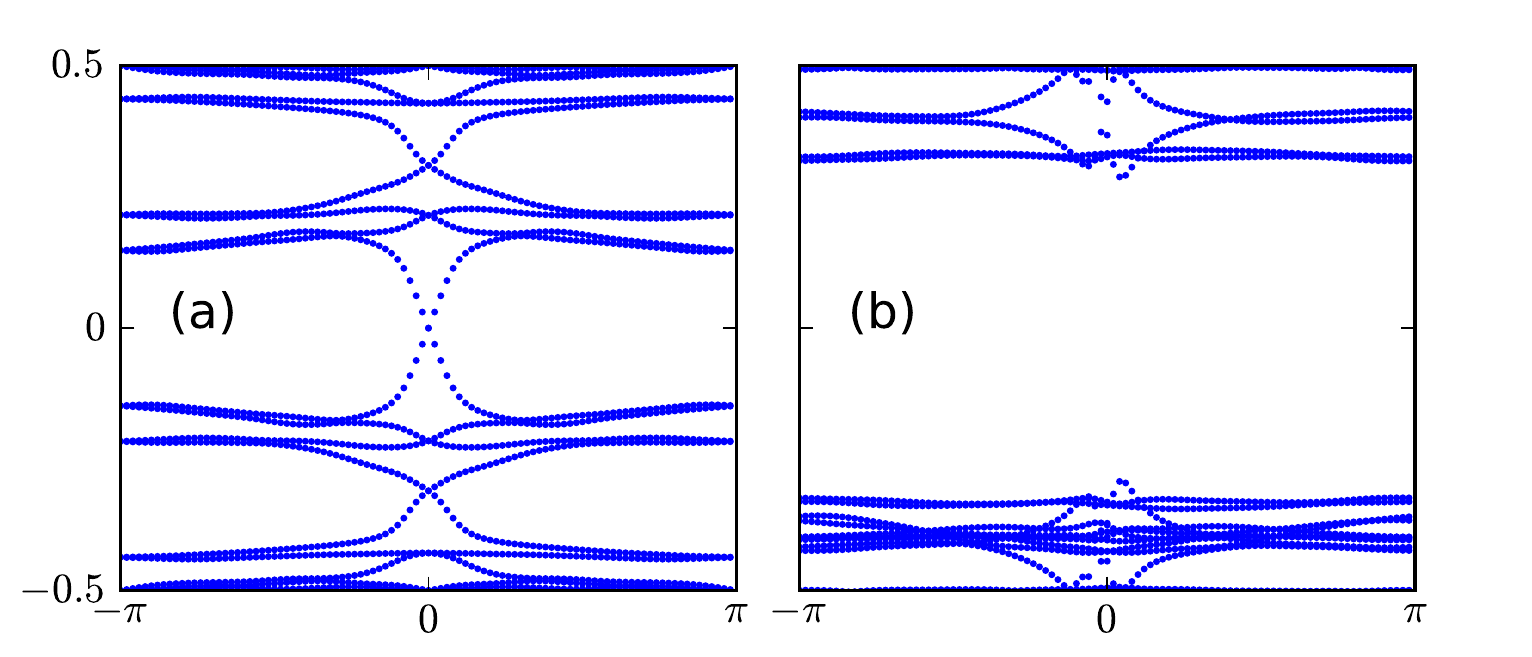}
  \caption{Flow of the hybrid Wannier centers in the $k_x=0$ plane in \bise{}. Panel (a): the gauge respects TR and inversion symmetries. Panel (b): the smooth gauge used to obtain the ELWFs. }
  \label{fig:wcc}  
\end{figure}

The Wannier spreads in the NI and TI phase are given in Tab.~\ref{tab:ws}. As expected we find them to be larger in the TI than in the NI phase.\cite{soluyanov_wannier_2011} Both symmetry breaking fields lead to nearly identical WFs, although, the resulting spread is found to be slightly larger when the field $\htr_\mathrm{BiSe}$ is used to break the symmetry. The smoothness of the resulting gauge is visible in Fig.~\ref{fig:wcc} (see Fig.~\ref{fig:trpol}), where we show the flow of the hybrid Wannier centers in the $k_x = 0$ plane. All centers are smooth and periodic in the BZ.

Now that a smooth gauge for \bise{} is obtained, the corresponding Bloch states can be used as a set of ${\bf k}$-dependent trial states $\tau_i({\bf k})$. These states are useful to quickly obtain a smooth gauge and ELWFs on a different $k$-mesh without repeating the above procedure. In the spirit of Wannier interpolation,~\cite{Souza-PRB01} the new $\tau_i({\bf k})$ are obtained by Fourier transforming the ELWFs on the new ${\bf k}$-mesh.
%

The obtained ELWFs also allowed us to find the local trial states that work for \bise{}, that result in well localized WFs for different ${\bf k}$-meshes.\footnote{A naive choice for the \unexpanded{$|\tau_i \rangle$} to be just the ELWFs truncated outside the home unit cell is equivalent to averaging the smooth Bloch states over the BZ. This can cause a problem during projection, since the volume of the region in the BZ, where the band ordering is inverted, is often much smaller than that of the region with normal ordering. Therefore, this choice may underrepresent the band character corresponding to the inverted band ordering in the trial states, so \unexpanded{$\det{S(\bk )}$} will become small when projecting onto them. We found that this problem can easily be solved by manually adding a contribution of the dominating character \unexpanded{$| \rho \rangle$} of the smooth Bloch states at the region of band inversion to the trial states, so that the actual trial states become \unexpanded{$|\overline \tau_i \rangle = 1/\sqrt{2} (| \tau_i \rangle + |\rho \rangle)$}. Table~\ref{tab:trial} lists the trial states obtained with this procedure.}
They are listed in Tab.~\ref{tab:trial}, where the site location (see Fig.~\ref{fig:crys_struct}), the orbital character and the spin direction are provided for each of the trial states. Apart from the two exclusions, the states come in time-reversal pairs listed in the same row of the table. The trial states have mostly Se-character, apart from the exclusions, which are listed in rows 4 and 7 of the table, and reflect the band inversion that occurs between the $p_x$ states of Se and Bi in \bise{} at the $\Gamma$-point, which can be seen in Fig.~\ref{fig:bands}(b).

As discussed above, \bise{} is an example of a double symmetry protection of the band topology, where both inversion and TR are broken in the smooth gauge. This breaking of the two symmetries is clearly seen in the flow of hybrid Wannier centers shown in Fig.~\ref{fig:wcc}. Figure~\ref{fig:wcc}(a) shows hybrid Wannier centers constructed from a symmetry-preserving gauge. The topological obstruction is evident. In contrast,  Fig.~\ref{fig:wcc}(b) illustrates the case of the smooth gauge obtained above. The smooth gauge Wannier centers are neither inversion, nor TR-symmetric, but are smooth and periodic in the BZ, corresponding to zero individual Chern numbers.
\begingroup
\squeezetable
\begin{table}
\begin{tabular}{c | c}
  $|\mathrm{Se0}, $p$_x , \ua_z \rangle$ & $|\mathrm{Se0}, $p$_x , \da_z \rangle$ \\
  $|\mathrm{Se0}, $p$_y , \ua_z \rangle$ & $|\mathrm{Se0}, $p$_y , \da_z \rangle$ \\
  $|\mathrm{Se0}, $p$_z , \ua_z \rangle$ & $|\mathrm{Se0}, $p$_z , \da_z \rangle$ \\
  \hline
  $\frac{1}{\sqrt 2}(|\mathrm{Se1}, $p$_x, \ua_x \rangle + | \mathrm{Bi1}, $p$_x, \ua_x \rangle)$ & $|\mathrm{Se1}, $p$_x, \da_x \rangle$ \\
  $|\mathrm{Se1}, $p$_y , \ua_x \rangle$ & $|\mathrm{Se1}, $p$_y , \da_x \rangle$ \\
  $|\mathrm{Se1}, $p$_z , \ua_x \rangle$ & $|\mathrm{Se1}, $p$_z , \da_x \rangle$ \\
  \hline
  $|\mathrm{Se2}, $p$_x, \ua_x \rangle$ &  $\frac{1}{\sqrt 2}(|\mathrm{Se2}, $p$_x, \da_x \rangle + | \mathrm{Bi2}, $p$_x, \da_x \rangle)$ \\
  $|\mathrm{Se2}, $p$_y , \ua_x \rangle$ & $|\mathrm{Se2}, $p$_y , \da_x \rangle$ \\
  $|\mathrm{Se2}, $p$_z , \ua_x \rangle$ & $|\mathrm{Se2}, $p$_z , \da_x \rangle$ 
\end{tabular}
\caption{The 18 trial states for the topological material \bise{}. Left and right column show potential time reversed partners.}
\label{tab:trial}
\end{table}
\endgroup
\subsection{Evaluation of the Chern-Simons magnetoelectric polarizability $\theta_\mathrm{CS}$}
\label{sec:bise_magneto}
We now proceed to calculating the geometrical contribution to the magnetoelectric effect.
The numerical evaluation of the Chern-Simons magnetoelectric polarizability $\theta_\mathrm{CS}$ introduced in Sec.~\ref{sec:intro} is a tedious task.\cite{Coh_Chern-Simons_2011} While a direct simulation of a material's response to electromagnetic fields can potentially be used to evaluate this term, such a calculation requires the use of large supercells, which makes it inefficient and computationally expensive. Several methods to compute $\theta_\mathrm{CS}$ from the bulk wave functions exist,~\cite{Essin_magnetoelectric_2009,essin_magnetoelectric_2010,Coh_Chern-Simons_2011,Taherinejad_adiabatic_2015,liu_theta} but only one of them (Ref.~\onlinecite{Coh_Chern-Simons_2011}) was applied to materials within an {\it ab initio} framework. Our calculation is based on the formalism developed in the latter work, but the actual implementation of the formulas is also done directly in position space, which  significantly improves  convergence with respect to the ${\bf k}$-mesh, compared to the ${\bf k}$-space implementations used before.\cite{Coh_Chern-Simons_2011}

To calculate $\theta_\mathrm{CS}$ the Bloch wave functions obtained from the bulk calculation are used. The Berry connection matrix for these states is defined  as
\begin{equation}
\A_{mn,j}(\mathbf k) = \langle u_{m\mathbf k} | i \frac{\partial}{\partial k_j} | u_{n\mathbf k} \rangle ,
\end{equation}
which is computed using the smooth gauge, obtained above.~\footnote{This is done using the smooth Bloch functions obtained for the projected Hamiltonian as \unexpanded{$\mathcal A_{mn,j}(\mathbf k) \simeq i \sum_{\bf b} w_{\bf b} b_j (\langle u_{m\mathbf k} | u_{n \mathbf{k} + \mathbf{b}} \rangle - \delta_{mn})$}, where $\mathbf b$ are the vectors connecting $\mathbf k$ to its nearest neighbors on the ${\bf k}$-mesh.
The overlap matrices \unexpanded{$\langle u_{m\mathbf k} | u_{n \mathbf{k} + \mathbf{b}} \rangle$} are 
calculated assuming the basis WF, used to construct the projected Hamiltonian, to be the eigenstates of the position operator. Certain (small) corrections that account for this approximation, and arise only when the TR-symmetry is broken, are omitted here. Their calculation, possible in principle,~\cite{Wang_Hall_2006} goes beyond the scope of the present paper.}
Then $\theta_\mathrm{CS}$ is evaluated by integrating the Chern-Simons 3-form over the entire Brillouin zone~\cite{Qi_topological_2008,Essin_magnetoelectric_2009}
\begin{equation}
  \theta_\mathrm{CS} = -\frac{1}{4\pi} \int_\mathrm{BZ} d^3 k\, \epsilon_{ijk} \mathrm{tr}\left[ \A_i \partial_j \A_k -\frac{2i}{3} \A_i \A_j \A_k \right],
  \label{eq:theta}
\end{equation}
where the summation over band and Cartesian indices is assumed. While this integral is gauge invariant modulo $2\pi$, a smooth gauge is assumed in the derivation of Eq.~\ref{eq:theta}.\cite{Qi_topological_2008,essin_magnetoelectric_2010} Thus a smooth gauge is required for a meaningful evaluation of the integral. This is when the above technique for finding a smooth gauge becomes important.

Equation~(\ref{eq:theta}) can be rewritten in terms of matrix elements of position operators evaluated with WFs in position space. Using
\begin{equation}
  A_{mn,j}( \mathbf k) = \sum_{\mathbf R} e^{i \mathbf k \cdot \mathbf R} \langle \mathbf 0 m | r_j | \mathbf R n \rangle,
  \label{eq:wme}
\end{equation}
the result reported previously\cite{Coh_Chern-Simons_2011} can be obtained
\begin{align}
    \theta_\mathrm{CS} = &\frac{1}{4\pi} \frac{2 \pi^3}{\Omega} \epsilon_{ijk}\, \mathrm{Im}\biggl( \sum_\mathbf{R} \langle \mathbf{0} m | r_i | \mathbf R n \rangle \langle \mathbf{R} n | r_i | \mathbf{0} m \rangle R_k \nonumber  \\
    &- \frac 2 3 \sum_{\mathbf{RP}} \langle \mathbf 0 l | r_i | \mathbf R m \rangle \langle \mathbf R m | r_j | \mathbf P n \rangle \langle \mathbf P n | r_k | \mathbf 0 l \rangle  \biggr).
    \label{eq:theta1}
\end{align}
This equation still assumes a smooth gauge, since when the WFs are not exponentially localized, surface terms should be included to account for the slow decay of the Wannier matrix elements.\citep{Coh_Chern-Simons_2011}
\begin{figure}
  \includegraphics[width =  0.45 \textwidth]{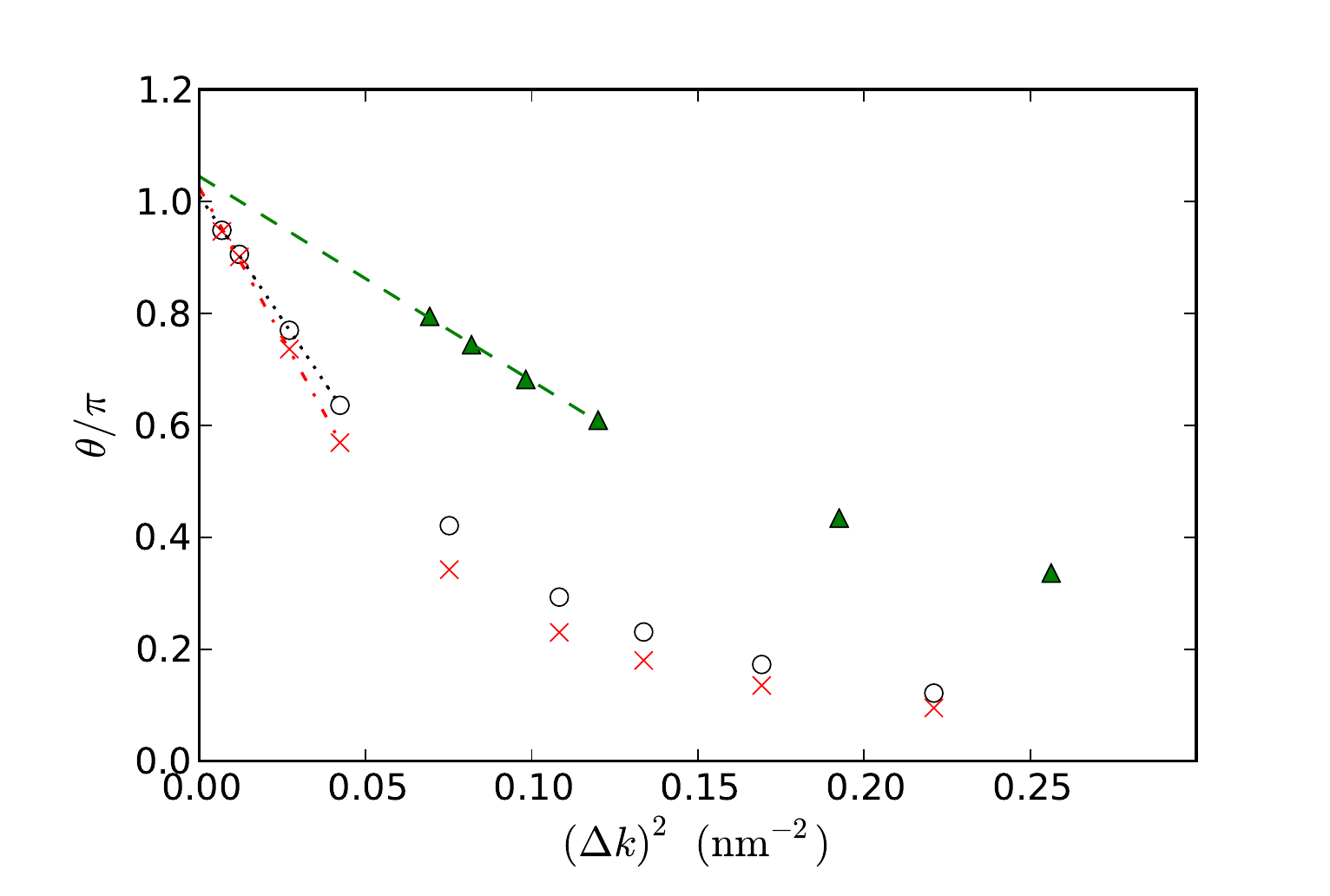}
  \caption{Convergence of $\theta_\mathrm{CS}$ in the TI for varying densities of ${\bf k}$-meshes. $\Delta k$ is the nearest-neighbor spacing on the grid. Convergence to $\theta_\mathrm{CS}=\pi$ can be reached for mesh densities of order 80$\times$80$\times$80 ($(\Delta k)^2=0.0068\,\mathrm{nm}^{-2}$). The black circles (red crosses) correspond to the results obtained by evaluating the Berry connection in ${\bf k}$-space without (with) maximal localization of the WFs resultant from the projections of Tab.~\ref{tab:trial}. The green triangles correspond to the results obtained by evaluating $\langle \mb 0 m | r_j | \mb R n \rangle$ in position space. A linear extrapolation to the infinitely dense ${\bf k}$-mesh was done using the last data points.}
  \label{fig:bisetheta}
\end{figure}

We now use the ELWFs obtained for \bise{} above, to compute the $\theta_\mathrm{CS}$-term in this material.
While both formulas Eq.~\eqref{eq:theta} and Eq.~\eqref{eq:theta1} agree in the limit of an infinitely dense ${\bf k}$-mesh, the use of the position space formula of Eq.~\eqref{eq:theta1} results in faster convergence, compared to the ${\bf k}$-space formulation of Eq.~\eqref{eq:theta}.~\footnote{
The first term of Eq.~\eqref{eq:theta} contains ${\bf k}$-derivatives $\partial_j \A_k$ that cause slow convergence. Using Eq.~\eqref{eq:wme} the derivatives can be treated analytically as it is done in Eq.~\eqref{eq:theta1}.  The convergence of the first term with respect to the ${\bf k}$-mesh is improved. The convergence of the second term in Eqs.~\ref{eq:theta}-\ref{eq:theta1} is found to be identical in both real and momentum space formulations. However, ${\bf k}$-space formulation of the Eq.~\eqref{eq:theta} was used for the evaluation of this term being more efficient numerically.} 

We used the trial states of Tab.~\ref{tab:trial} to get localized WFs for a variety of ${\bf k}$-meshes up to 80$\times$80$\times$80. The scaling in the TI phase is shown in Fig.~\ref{fig:bisetheta}. Because of the slow convergence in ${\bf k}$-mesh density, an extrapolation to the infinitely dense mesh is required. A linear extrapolation in $(\Delta k)^2$ is done using the last data points, getting in all cases very close to the expected value of $\theta_\mathrm{CS} = \pi$. This extrapolation choice is dictated by the error in numerical evaluation of  the $\mathcal A_j$ matrices and the Wannier matrix elements $ \langle \mathbf 0 m | r_j | \mathbf R n \rangle$, which is of order ${\cal O}((\Delta k)^2)$.\cite{Wang_Hall_2006}

Exponential convergence can be achieved by directly evaluating $\langle \mb 0 m | r_j | \mb R n \rangle$ in position space with the ELWFs.\cite{stengel_wannier_2006} This is, however, computationally more expensive and was done only for relatively coarse ${\bf k}$-meshes up to 25$\times$25$\times$25. The green triangles in Fig.~\ref{fig:bisetheta} illustrate the results obtained from this position space approach. Even with this coarse ${\bf k}$-mesh a reliable extrapolation to the expected value $\theta_\mathrm{CS} = \pi$ can be done within a few percent accuracy.

\begin{figure}
  \includegraphics[width =  0.45 \textwidth]{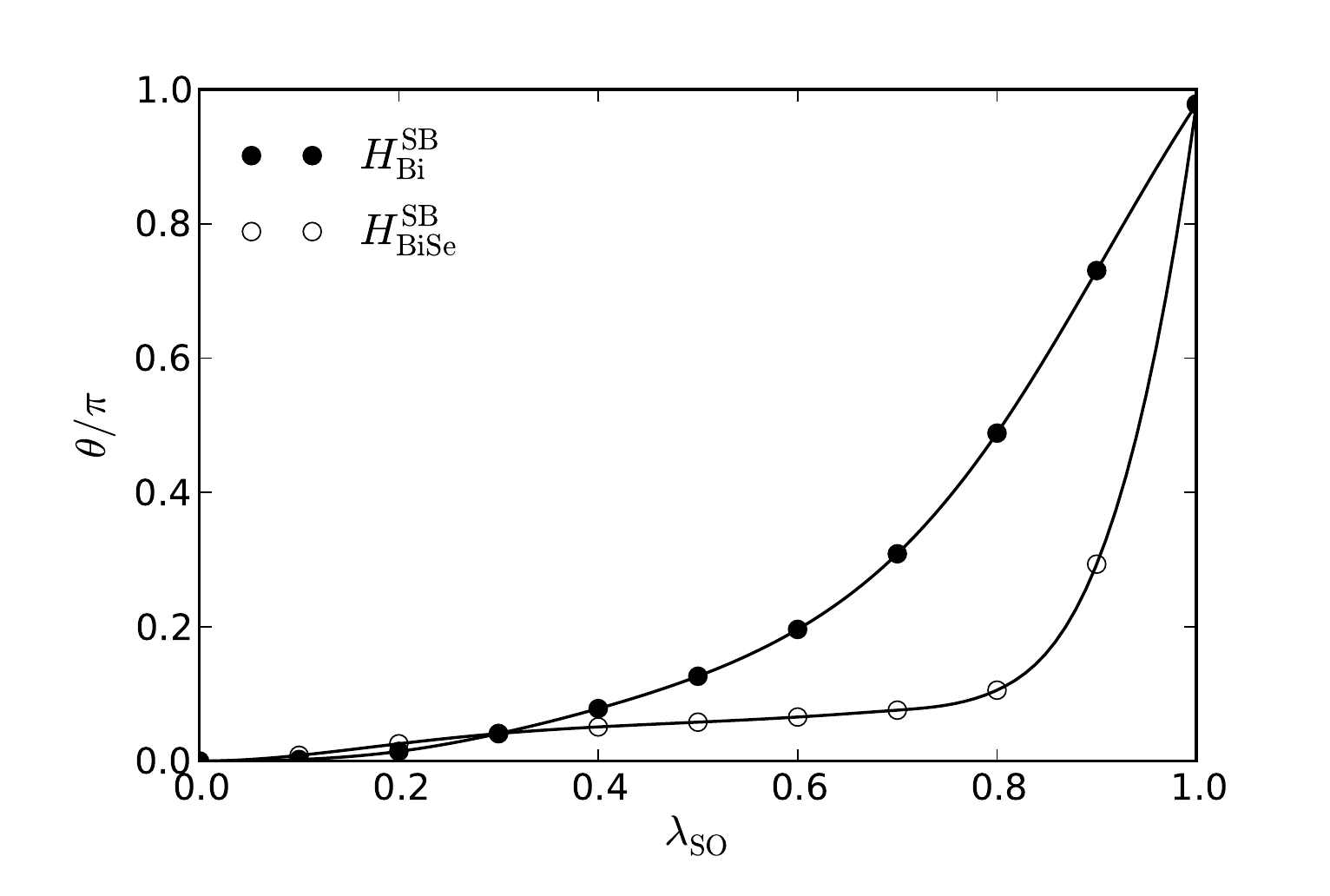}
  \caption{$\theta_\mathrm{CS}$ calculated along the adiabatic path connecting NI and TI phases. The full (empty) circles correspond to the symmetry breaking field $\htr_\mathrm{Bi}$ ($\htr_\mathrm{BiSe}$).\cite{Note5}}
  \label{fig:bisetheta_path}
\end{figure}

The $\theta_\mathrm{CS}$-term was also computed along the adiabatic path connecting NI to TI for the two different symmetry breaking fields ($\htr_\mathrm{Bi}$ and $\htr_\mathrm{BiSe}$). The results are shown in Fig.~\ref{fig:bisetheta_path}. The $\theta_\mathrm{CS}$-term is found to be larger for the $\htr_\mathrm{Bi}$ field, which does not break the product symmetry of TR and inversion.
\section{Conclusion}
\label{sec:concl}
In this paper we established a general procedure for constructing ELWFs and smooth Bloch states to describe a group of bands with nontrivial topology. This is done by connecting the topologically nontrivial Hamiltonian to some trivial one by a gapped adiabatic path. Our technique works for all symmetry-protected topologies, provided that the net Chern number of the bands is zero. It was illustrated for $\zz$ TIs in two dimensions using the example of the Kane-Mele tight-binding model, and the real three-dimensional topological insulator material \bise{}. We also introduced the concept of double symmetry protection of nontrivial topology, which describes (ubiquitous) situations when there is more than one symmetry that presents the obstruction for choosing smooth Bloch states. We expect that this technique can also be generalized to the case of Chern insulators to obtain numerically a smooth lattice representation for all the occupied Bloch states, apart from one that carries the net Chern number of the system, and hence cannot be made smooth.

Finally, we described how the proposed scheme of constructing a smooth gauge allows for the calculation of the Chern-Simons $\theta_\mathrm{CS}$-term that captures the geometric contribution to the orbital magnetoelectric response of materials. A detailed discussion of the numerical implementation of this calculation was provided, showing how to efficiently implement our technique to materials, where $\theta_\mathrm{CS}$ is not quantized. This technique is especially useful in the presence of band topologies that do not result in a quantized value of the $\theta_\mathrm{CS}$-term. More generally, the numerical construction of the smooth gauge for lattice models presented here can have broader applications in evaluation of various (band) geometric effects that do not have an immediate gauge-invariant formulation.

\section{Acknowledgments}
We would like to thank David Vanderbilt and Xiao-Liang Qi for useful discussions. This work was supported by Microsoft Research, the European Research Council through ERC Advanced Grant SIMCOFE, and the Swiss National Science Foundation through the National Competence Centers in Research MARVEL and QSIT.
\appendix*
\section{Symmetries of the Kane-Mele model}
\label{appendix}
Here the symmetries of the Kane-Mele model discussed in the main text are derived explicitly. Using the basis set $\left( |A \uparrow \rangle , |B \uparrow \rangle , |A \downarrow \rangle , |B \downarrow \rangle  \right) $ the Hamiltonian Eq.~\eqref{eq:kmH} can be written in ${\bf k}$-space
\begin{equation}
 \begin{split}
H(\mathbf k) =\  & t\, \sigma_0 [\tau_x \epsilon_{t1}(\mathbf k) + \tau_y \epsilon_{t2}(\mathbf k) ] \\
     + &\lso\, \sigma_z \tau_z \epsilon_\mathrm{SO}(\mathbf k) \\
     + &\lr\, \{ \sigma_x [  \tau_x \epsilon_{\mathrm{R}1}(\mathbf k) +  \tau_y \epsilon_{\mathrm{R}2}(\mathbf k)] \\
     &\ \  + \sigma_y [  \tau_y \epsilon_{\mathrm{R}3}(\mathbf k) +  \tau_x \epsilon_{\mathrm{R}4}(\mathbf k)] \} \\
     - &\lnu\, \sigma_0 \tau_z ,
  \end{split}
\end{equation}
with $\sigma$ and $\tau$ matrices acting in spin and lattice subspaces; $\sigma_0$ being the unit matrix, and $\sigma_{x,y,z}$ ($\tau_{x,y,z}$) -- the Pauli matrices. 

The $\mathbf k$-dependent coefficients in the Hamiltonian are
\begin{equation}
  \begin{split}
    \epsilon_{t1}(\mathbf k) & = (1+\delta) f_1(\mathbf k) + f_2(\mathbf k) + f_3(\mathbf k) ,\\
    \epsilon_{t2}(\mathbf k) & = (1+\delta) g_1(\mathbf k) + g_2(\mathbf k) + g_3(\mathbf k) ,\\
    \epsilon_\mathrm{SO}(\mathbf k) & = 2[-\sin(k_1) + \sin(k_2) + \sin(k_1-k_2)] ,\\
    \epsilon_{\mathrm{R}1} & = -\frac{1}{2} g_1(\mathbf k) - \frac{1}{2} g_2(\mathbf k) + g_3(\mathbf k) ,\\
    \epsilon_{\mathrm{R}2} & = -\frac{1}{2} f_1(\mathbf k) - \frac{1}{2} f_2(\mathbf k) + f_3(\mathbf k) ,\\
    \epsilon_{\mathrm{R}3} & = \frac{\sqrt{3}}{2} f_1(\mathbf k) - \frac{\sqrt{3}}{2} f_2(\mathbf k) ,\\
    \epsilon_{\mathrm{R}4} & = \frac{\sqrt{3}}{2} g_1(\mathbf k) - \frac{\sqrt{3}}{2} g_2(\mathbf k) ,\\
  \end{split}
\end{equation}
where 
\begin{equation*}
\begin{array}{ll}
  f_1(\mathbf k)  =   \cos(\frac{k_1}{3}+\frac{k_2}{3}), & g_1(\mathbf k)  = \sin(\frac{k_1}{3}+\frac{k_2}{3}), \\
  f_2(\mathbf k)  =  \cos(-2\frac{k_1}{3}+\frac{k_2}{3}), & g_2(\mathbf k)  = \sin(-2\frac{k_1}{3}+\frac{k_2}{3}), \\
  f_3(\mathbf k)  = \cos(\frac{k_1}{3}-2\frac{k_2}{3}), & g_3(\mathbf k)  =  \sin(\frac{k_1}{3}-2\frac{k_2}{3}),
\end{array}
\end{equation*}
with $k_1$ and $k_2$ being the reduced coordinates in terms of the reciprocal lattice vectors $\boldsymbol{b}_1=\frac{2\pi}{a}(\boldsymbol{\hat k}_x - \frac{1}{\sqrt 3} \boldsymbol{\hat k}_y)$ and $\boldsymbol{b}_2=\frac{2\pi}{a}\frac{2}{\sqrt 3} \boldsymbol{\hat  k}_y$.

\begin{table}[t]
\begin{tabular}{l | c | c | c}
  symmetry    & representation $\hat R$ & $k_1 \rightarrow \dots$ & $k_2 \rightarrow \dots$ \\
  \hline 
  $\sigma_\nu^{(11)}$ & $i\left( -\frac{\sigma_x}{2}+\frac{\sqrt{3}}{2}\sigma_y \right)  \tau_0$ & $k_2 $  & $ k_1 $ \\
  $\sigma_\nu^{(1\overline 1)}$ & $i\left( \frac{\sqrt{3}}{2}\sigma_x+\frac{\sigma_y}{2} \right)\tau_x$ & $ -k_2 $ & $-k_1 $ \\
    $C_2$ & $i \sigma_z \tau_x$ & $ -k_1 $ & $-k_2 $ \\
  $C_3$ & $\left( \frac{\sigma_0}{2} + i \frac{\sqrt{3}}{2} \sigma_z \right) \tau_0$ & $ k_2-k_1$ & $ -k_1 $ \\
  $C_6$ & $\left( \frac{\sqrt{3}}{2} \sigma_0 + i  \frac{\sigma_z}{2} \right) \tau_x$ & $ k_2$ & $ k_2-k_1 $ \\
\end{tabular}
\caption{Point symmetries (relative to the origin) of the KM model for $\lnu = \delta = 0$. Both momentum transformations and matrix representations are given.}
\label{tab:sym}
\end{table}

We first consider the case of $\lnu = \delta = 0$, corresponding to the point P3 in Fig.~\ref{fig:km_pd}. In this case the 2D space group (wallpaper group) of the model is $p6m$ in IUC notation (*632 in orbifold notation),~\cite{conway} and the corresponding 2D point group is $\mathrm{D}_6$. The specific matrix representation of all the point group symmetries for this case is provided in Tab.~\ref{tab:sym}.
Each of the symmetries acts according to
\begin{equation}
H(S(k_1,k_2))=\hat R_{S} H(k_1, k_2) \hat R_{S}^\dag ,
\end{equation}
where $S$ is the symmetry operation and $\hat{R}_S$ is its matrix representation.

The consistency of the matrix representation of the Tab.~\ref{tab:sym} can be checked by noting that the $C_2$ symmetry is the product of the two mirrors $\sigma_\nu^{(11)}$ and $\sigma_\nu^{(1\overline 1)}$, and the $C_3$ symmetry is the $C_6$ rotation applied twice, as expected. Other reflection symmetries can be obtained by appropriate combinations of these operations. The relations expected for spin-$1/2$ systems, namely $\sigma^2=C_2^2=C_3^3=C_6^6=-1$ are satisfied. Without the Rashba term, that is for $\lr = 0$, there would be an additional inversion symmetry $\sigma_0 \tau_x$.

The symmetries corresponding to the different points in the path of Fig.~\ref{fig:km_pd} are summarized in Tab.~\ref{tab:sym1}. When both $\lnu\neq 0$ and $\delta\neq 0$, which is the case on the line connecting $\mathrm{P1}$ and $\mathrm{P2}$ ($\overline{\mathrm{P1}\,\mathrm{P2}}$) in Fig.~\ref{fig:km_pd}), the sum of the Wannier centers $\boldsymbol P$ is allowed to change continuously along the lines $P_y = \frac{1 }{\sqrt 3} P_x + \frac m 2$, where $m$ is an integer.
\begin{table*}
\begin{tabular}{ l | c | c | c | c}
  & $\lnu=\delta=0$ &  $\lnu\ne 0, \delta =0$ &  $\lnu = 0, \delta \ne 0$ &  $\lnu \ne 0, \delta\ne 0$ \\
  \hline
  see Fig.~\ref{fig:km_pd} & P3 & P1 & $\overline{\mathrm{P2}\,\mathrm{P3}}$ & $\overline{\mathrm{P1}\,\mathrm{P2}}$ \\
  wallpaper group & p6m, *632 & p3m1, *333 & cmm, 2*22 & cm, *x \\
  point group  & D$_6$  & D$_3$  & D$_2$  & D$_1$ \\
  conserved symmetries & $\sigma_\nu^{(11)},\sigma_\nu^{(1\overline 1)},C_2,C_3,C_6$ & $\sigma_\nu^{(11)},C_3$ & $\sigma_\nu^{(11)},\sigma_\nu^{(1\overline 1)},C_2$ & $\sigma_\nu^{(11)}$ \\
  $\boldsymbol{P}$ quantized & Yes & Yes & Yes & No \\
\end{tabular}
\caption{Symmetries at the different points of the adiabatic path of Fig.~\ref{fig:km_pd}. The wallpaper group is given in IUC and orbifold notation}
\label{tab:sym1}
\end{table*}

The TR symmetry in the model is given by $i \sigma_y \tau_0 \mathcal{K}$, with $\mathcal{K}$ being complex conjugation, and its action in ${\bf k}$-space given by $k_1 \rightarrow -k_1$ and $k_2 \rightarrow -k_2$.
Two TR-breaking fields, $\htr_{(1)}$ and $\htr_{(2)}$ were considered in this work
\begin{equation*}
  \begin{split}
    \htr_{(1)}  & = (-\frac{1}{2} \sigma_x + \frac{\sqrt 3}{2}\sigma_y)\tau_z, \\
    \htr_{(2)} & = (-\frac{\sqrt 3}{4} \sigma_x + \frac{3}{4}\sigma_y+\frac{1}{2} \sigma_z)\tau_z .
  \end{split}
\end{equation*}
Of them, $\htr_{(1)}$ commutes with both mirror symmetries $\sigma_\nu^{(11)}$,  $\sigma_\nu^{(1\overline 1)}$. The field is applied along the line $\overline{\mathrm{P_2}\,\mathrm{P_3}}$, where, according to Tab.~\ref{tab:sym1}, both these mirrors are preserved. Therefore, the resulting WFs reflect this symmetry and the minimization of their spread gets stuck in a local minimum. By adding the $\sigma_z$ component, leading to $\htr_{(2)}$, both mirrors are broken and we obtain the maximally localized WFs also found in Ref.~\onlinecite{soluyanov_wannier_2011}.
%

\bibliography{literature}
\end{document}